\documentclass[usegraphicx,usenatbib,natbib,useAMS]{mn2e}

\usepackage{times}     
\usepackage{aas_macros}
\usepackage{amssymb}   
\usepackage{multirow}

\def\ebv{\mbox{$E(B-V)$}}
\def\halpha{\mbox{H$\alpha$}}
\def\hbeta{\mbox{H$\beta$}}
\def\hgamma{\mbox{H$\gamma$}}
\def\hdelta{\mbox{H$\delta$}}
\def\lya{\mbox{Ly$\alpha$}}

\def\ecs{\mbox{\,erg~cm$^{-2}$~s$^{-1}$}}

\def\lesssim{\mathrel{\hbox{\rlap{\hbox{\lower4pt\hbox{$\sim$}}}\hbox{$<$}}}}
\def\gtrsim{\mathrel{\hbox{\rlap{\hbox{\lower4pt\hbox{$\sim$}}}\hbox{$>$}}}}

\DeclareRobustCommand{\ion}[2]{%
\relax\ifmmode
 \ifx\testbx\f@series
  {\mathbf{#1\,\mathsc{#2}}}\else
  {\mathrm{#1\,\mathsc{#2}}}\fi
 \else\textup{#1\,{\mdseries\textsc{#2}}}%
\fi}

\title[Low Mass Lensed Galaxies]{The Low Mass End of the Fundamental
  Relation for Gravitationally Lensed Star Forming Galaxies at $1<z<6$
 \thanks{Based on data from the
    X-shooter GTO observations collected at the European Southern
    Observatory VLT/Kuyuen telescope, Paranal, Chile, under programme
    IDs: 084.B-0351(D), 086.A-0674(A), 086.A-0674(B), 087.A-0432(A)
    and 087.A-0432(B). Based on $\emph{HST}$ general observer
    programmes GO-10491, GO-11103, and GO-12166.}}

\author[L. Christensen et al.]
       {Lise Christensen\thanks{lise@dark-cosmology.dk}$^{1,2}$,
         Johan Richard$^{3,1}$, 
         Jens Hjorth$^1$, 
         Bo Milvang-Jensen$^1$, 
         \newauthor 
         Peter Laursen$^{4,1}$, 
         Marceau Limousin$^5$,
         Miroslava Dessauges-Zavadsky$^6$, 
         \newauthor
         Claudio Grillo$^{1,2}$, 
         Harald Ebeling$^7$\\
        $^1$ Dark Cosmology Centre, Niels Bohr Institute, University of Copenhagen, Juliane Maries Vej 30, 2100 Copenhagen, Denmark\\
        $^2$ Excellence Cluster Universe, Technische Universit\"{a}t
        M\"{u}nchen, Bolzmanstrasse 2, 85748 Garching, Germany\\
        $^3$ Centre de Recherche Astrophysique de Lyon, Universit´e Lyon 1, Observatoire de Lyon, 9 Avenue Charles Andr´e, 69561 Saint Genis
Laval cedex, France\\
        $^4$ Oskar Klein Centre, Dept. of Astronomy, Stockholm University, SE-10691 AlbaNova, Stockholm, Sweden\\
        $^5$ Aix Marseille Universit\'e, CNRS, LAM (Laboratoire d'Astrophysique de Marseille) UMR 7326, 13388, Marseille, France\\
        $^6$ Observatoire de Gen\`eve, Universit\'e de Gen\`eve, 51 Ch. des Maillettes, 1290 Sauverny, Switzerland\\
        $^7$ Institute for Astronomy, University of Hawaii, 2680 Woodlawn Drive, Honolulu, HI 96822, USA\\
}
\date{Accepted 2012 August 29. Received 2012 August 28}      
\pubyear{2012}

\begin{document}

\maketitle

\label{firstpage}

\begin{abstract}
We present VLT/X-shooter spectra of 13 galaxies in the redshift range
$1\lesssim z\lesssim6$, which are strongly lensed by massive galaxy
clusters. Spectroscopic redshifts are measured for nine galaxies,
while three sources have redshifts determined from continuum breaks in
their spectra. The stellar masses of the galaxies span four orders of
magnitude between $10^7$ and $10^{11}$ M$_{\odot}$ and have
luminosities at 1500~{\AA} rest-frame between 0.004 and 9 $L^*$ after
correcting for the magnification. This allows us to probe a variety of
galaxy types from young, low-mass starburst galaxies to massive
evolved galaxies. The lensed galaxies with stellar masses less than
$10^{10}$M$_{\odot}$ have a large scatter compared to the fundamental
relation between stellar mass, star formation rates and oxygen
abundances. We provide a modified fit to the fundamental relation for
low-mass, low-metallicity galaxies with a weaker dependence of the
metallicity on either the star formation rate or stellar mass compared
to low-redshift, high-mass and high-metallicity SDSS galaxies.
\end{abstract}


\section{Introduction}

Strong gravitational lensing is an important tool for characterising
the dark matter distribution and formation of large mass
concentrations in a $\Lambda$CDM framework. Since massive galaxy
clusters give rise to magnification factors of 10--50 of background
galaxies, as well as stretching the source images into extended arcs,
lensing can be used to characterise the physical properties of
intrinsically sub-luminous galaxies and investigate spatially resolved
properties of high-redshift galaxies \citep{swinbank07}. Gravitational
lensing is also valuable for locating $z>6$ galaxies
\citep{bradley08,richard11b,zitrin12} and investigate individual
galaxies which may be representative of those responsible for
reionisation.

The Lyman break technique by construction selects galaxies with
similar characteristics, such that their stacked spectra can be
analysed \citep{shapley03}. Individual galaxies are too faint for
spectroscopic studies with reasonable use of telescope time, apart
from unusually bright Lyman break galaxies (LBGs) \citep{erb10}.
Other selection techniques have been used to study the fainter end of
the high redshift galaxy luminosity function, including \lya\ emitters
\citep[e.g.][]{fynbo03,nilsson11} or host galaxies of gamma-ray bursts
\citep[e.g.][]{hjorth12}. However, they fail to reach the same level
of detail because of the large amount of telescope time needed. Our
current understanding of the physical properties of high-redshift
galaxies is therefore limited to a small range in absolute magnitude.

Gravitational lensing allows us to study intrinsically fainter
galaxies with unusual spectral characteristics at any given redshift,
although most lensed galaxies studied in detail to date still belong
to the high-mass, high-luminosity end.  A few cases have provided a
great deal of insight into the physical properties of individual
$z=2$-$3$ galaxies, with the prime example being the lensed galaxy
MS~1512-cB58 \citep{pettini00,teplitz00,pettini02}.  The past few
years have seen a steady increase in the number of spectroscopic
observations of lensed galaxies at $z\sim2$ either detected
serendipitously or in large dedicated surveys
\citep{fosbury03,cabanac05,swinbank07,allam07,stark08,lin09,diehl09,bian10,rigby11,richard11,wuyts12}. Whereas
some $z=2$ lensed galaxies are easily investigated with low-resolution
spectra from 8-m class telescopes, few are sufficiently bright for
medium- or high-resolution spectroscopy, which is necessary for
examining the interstellar medium of the galaxy.  Rest-frame UV
absorption line spectra allow a detailed analysis of the interstellar
medium of galaxies with intrinsically very high star formation rates
\citep{cabanac08,quider09,quider10,dessauges10}.

As the redshift around $z\sim2$ represent the era of the peak of the
star formation, the lensed galaxies are relevant to compare to scaling
relations found in other galaxy samples, such as mass-metallicity
relations \citep{tremonti04}. Recently, a fundamental relation between
the mass, metallicity and SFR has been found. The relation has been
calibrated for metal-rich (oxygen abundances 12+log(O/H)$>$8.2), and
relatively massive galaxies (stellar masses above $10^{9.2}$
M$_{\odot}$) at $z<0.1$ \citep{lara-lopez10}. At higher redshifts up
to $z\approx2.5$ and for galaxies with lower metallicities the
relation appears to still hold \citep{mannucci10}. Strong
gravitational lensing of intrinsically fainter and less massive
galaxies allows us to explore the fundamental relation at even lower
masses and metallicities.  \citet{richard11} and \citet{wuyts12}
analysed the rest-frame optical emission lines of lensed galaxies at
$z=1-3$ with intrinsic low luminosities (0.1~$L^*$) and showed that
the galaxies are more metal-rich than predicted from extrapolations of
mass-metallicity relations.

In this paper, we present shallow spectroscopic observations from
VLT/X-shooter of 13 lensed galaxies selected towards massive galaxy
clusters at $z\sim0.4$. Section~\ref{sect:data} presents the data and
Section~\ref{sect:results} notes on each target and their redshift
determination. In Section \ref{sect:specprop} we explore the galaxy
properties from their continuum while in Section~\ref{sect:emphys} we
derive physical properties based on their emission lines.  In
Section~\ref{sect:fmr} we present an extension of the calibration at
the low-mass, low-metallicity end of the fundamental relation for star
forming galaxies.  Section~\ref{sect:conclusions} presents a summary.

\section{Observational data}
\label{sect:data}
One of the science goals for the 2$^{\mathrm{nd}}$ generation VLT
instrument X-shooter is to determine redshifts for faint single
targets and analyse emission line galaxies at $z>1$.  X-shooter is a
multi-wavelength cross-dispersed echelle spectrograph which splits up
the light by dichroics into three arms: UV-blue (UVB), which covers
3000--5600 {\AA}, visible-R (VIS), which covers 5600--10000 {\AA}, and
near-IR (NIR), which covers from 1 to 2.5 micron
\citep{dodorico06,vernet11}.  The slit length of X-shooter is
11\arcsec\ and the chosen slit widths determine the spectral
resolution. For the observations the slit widths were set to match the
optical seeing measured at the telescope as listed in
Table~\ref{tab:log}. Slit widths of 1.0/0.9/0.9 arcsec in the UVB,
VIS, and near-IR arms, give spectral resolutions of 5100, 8800, and
5600 respectively, while for the 1.3/1.2/1.2 arcsec slit widths, the
resolutions are 4000, 6700, and 4300. For all the observations, the
UVB CCD was read out binned by a factor of two in the dispersion
direction.

The instrument has proven to be useful for the study of gravitational
lensed galaxies due to its complete wavelength coverage and high
sensitivity throughout the spectral range. Physical properties of
lensed galaxies can be derived from short exposures as demonstrated
from commissioning data \citep{dessauges10,pettini10,christensen10}.
A wealth of information of the physical properties of galaxies can be
gained from the full medium-resolution UV- to near-IR spectral
coverage for both lensed galaxies such as the 8 o'clock arc
\citep{dessauges10,dessauges11} or even fainter gamma-ray burst (GRB)
host galaxies \citep{kruhler12}.  With the large wavelength coverage,
X-shooter is also ideal to determine faint galaxy redshifts, which is
necessary to construct the lens models \citep[e.g.][]{grillo11}.  Due
to the simultaneous coverage, there are no random offsets in the
placement of the slit between the three arms, which could be a problem
for optical spectra and near-IR spectra obtained with different
instruments. Such offsets could introduce differences in the flux
levels between the UV and near-IR galaxy spectra which are difficult
to correct for unless the continuum emission is well detected and can
be compared with broad-band photometry. In addition, systematic errors
in wavelength calibrations could be interpreted as kinematic effects
such as outflows. X-shooter is therefore ideal for the combined
purpose of determining redshift for faint, lensed sources, and allow a
characterisation of the physical nature of the lens galaxies.

Accurate mass models of the lensing clusters are required in order to
turn them into well-calibrated gravitational telescopes because of the
degeneracy between the source redshift and lens mass.  To determine
redshifts of lensed galaxies does not require high signal-to-noise
ratio spectra, and since the UV- selected high-redshift galaxies are
expected to be star-forming, and therefore have emission lines that
are much stronger than the underlying continuum, a redshift
determination can be obtained even for faint ($m=25$) galaxies with a
short integration time. With fairly short exposures, we can pinpoint
interesting targets for deeper follow-up studies.

\subsection{Target selection}
Some of the best calibrated cluster lenses are the $z<0.3$ Abell
clusters.  In this paper, we present spectra of galaxies without
previously determined spectroscopic redshifts lensed by Abell 1689,
where we aim to determine spectroscopic redshift for galaxies which
previously only had photometric redshifts. Partly, we focus on the
second large-scale dark matter concentration in the North-East
quadrant of the cluster \citep{limousin07}, such that the galaxy scale
mass substructure can be better constrained in this region.

We also choose to study galaxies lensed by higher redshift clusters.
We select targets belonging to the MAssive Cluster Survey (MACS),
which are massive X-ray luminous clusters at $0.3<z<0.7$
\citep{ebeling01}, and include observations of the southern equivalent
to the MACS with declinations $<$~--40 deg (H. Ebeling,
unpublished). These targets are denoted with the prefix SMACS.  Some
of the targets are also included in the CLASH survey
\citep{postman11}, allowing us to investigate in detail galaxies for
which high-quality, high-resolution images are available. Although the
16-broad band filter imaging allows detailed investigations of the
spectral energy distribution of the galaxies and their spatially
resolved properties, more accurate information can be obtained by
spectroscopic follow-up observations. The target galaxies are selected
from having high total brightness, and from those we investigate
regions with the highest surface brightness. This implies that we
preferentially select UV bright galaxies that are expected to show
significant emission lines, because the high surface brightness
regions are \ion{H}{ii} regions. Thereby, we can most easily determine
redshifts for the faintest galaxies, and also investigate the
kinematics and metal abundances in the emission line regions.

\begin{figure*}
\begin{center}
\includegraphics[width=17cm]{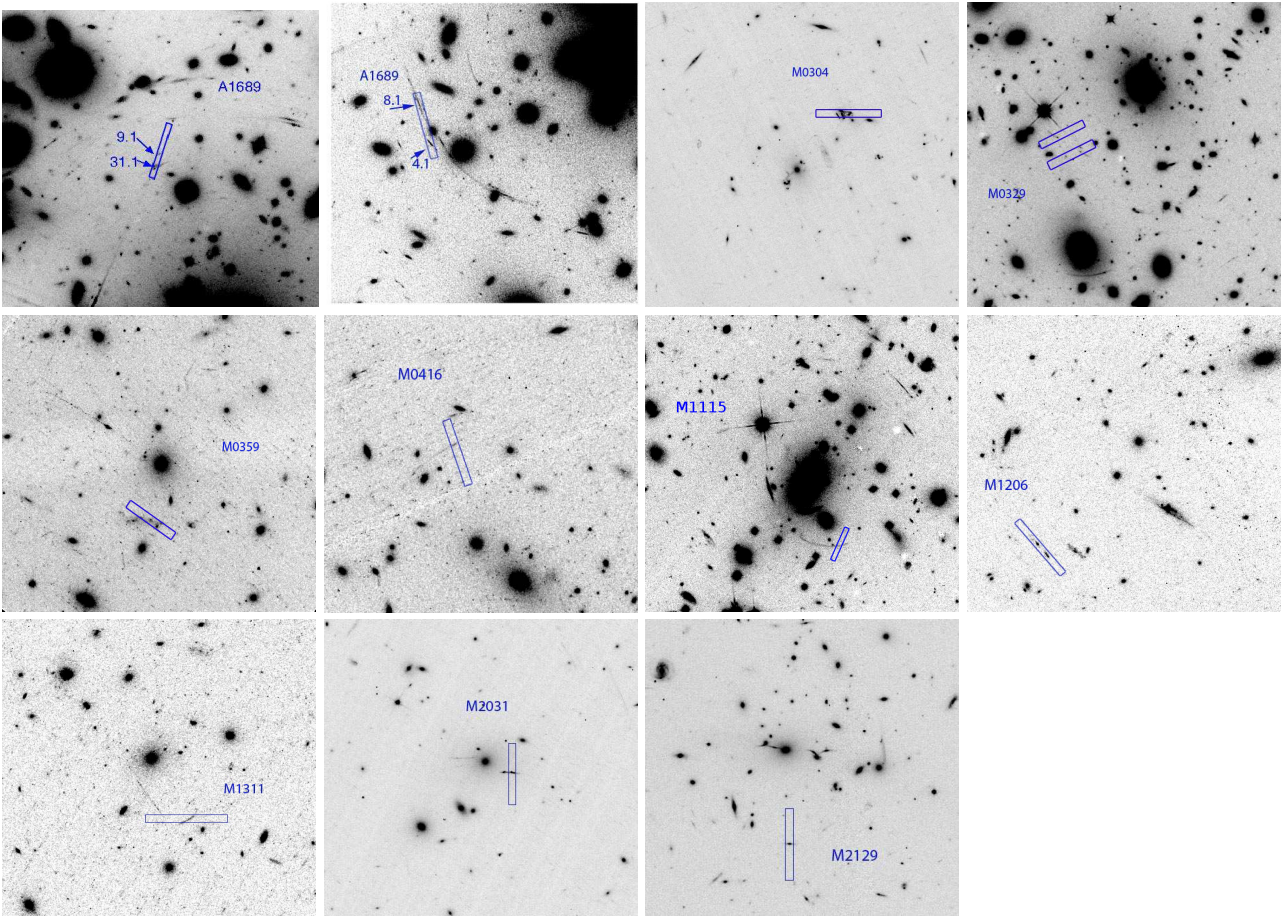}
\end{center}
\caption{Thumbnail \emph{HST} images of the clusters around the
  targeted sources. The images were taken with ACS, WFPC2 or WFC3 with
  either the F606W, F814W or F110W filters, and obtained through the
  MAST or HLA data archive {(\tt http://hla.stsci.edu/)}.  The images
  have varying sizes between 40 and 90 arcsec on a side with the
  orientation North up and East left, and with the X-shooter
  11\arcsec\ slit position indicated.}
\label{fig:piccy}
\end{figure*}

\subsection{Spectroscopic observations and data reduction}
The observations were done as a part of the X-shooter consortium
guaranteed observing time during several different runs in
2010--2011. A log of the observations can be found in
Table~\ref{tab:log}. The names of the arcs targeted for the
spectroscopic analysis are hereafter shortened, such that the names
reflect the cluster name only for the MACS clusters, i.e. the arc
observed in SMACS J0304.3–4402 is named M0304, and the arcs in Abell
1689 are denoted 'A' plus by their IDs, e.g. A4.1. In
Table~\ref{tab:log} and hereafter we adopt the image IDs from
\citet{limousin07} for Abell 1689 images, IDs for M0329 in
\citet{zitrin12}, IDs for M1206 from \citet{zitrin11b}, and the ID for
M2129 in \citet{zitrin11a}.

For target acquisition, relative offsets from bright stars were used,
so the location of the target on the slit depends on the accuracy of
the known star coordinates relative to the targets. The slit position
angle was not preferentially chosen along the long axis of the
extended arcs. In most cases, we chose a perpendicular angle such that
nodding the target along the slit would improve the residuals from sky
emission line subtraction.  Further, the angle was chosen to avoid
nearby objects.  The orientations of the slits are overlayed on the
panels in Fig.~\ref{fig:piccy}. A different procedure was used during
the first run, when the Abell 1689 cluster arcs were observed, where
we opted for longer time on target in stare-mode and did not move the
target along the slit. In addition, two arcs from different sources
were inside the slit. Whereas this procedure did allow us to determine
the arc redshifts, the quality of the near-IR spectra after the data
reduction was dominated by errors in sky subtraction, in particular
near sky emission lines. No separate sky exposures were obtained.

The data was reduced with the ESO pipeline version 1.3.7
\citep{goldoni06,modigliani10} using the physical model reduction
chain \citep{bristow08} in both nodding and staring mode.  The data
reduction works in a similar way for the three arms. First, the bias
level is subtracted from the UVB and VIS data (dark current in the
case of the near-IR data). The position of the orders on the detectors
were traced and a two-dimensional flat field was created. Then a
two-dimensional wavelength solution was determined from calibration
frames taken the day after the observations.  Cosmic rays were
rejected from the science frames using the LA Cosmic procedure
\citep{vandokkum01} within the pipeline.  Finally, the echelle orders
were extracted, rectified and merged onto a two-dimensional spectrum
using the errors as weights for the region of the overlap between the
different orders.

\begin{table*}
    \caption{Observing log}
  \begin{tabular}{lcrrcclc}
\hline
\hline
Cluster &  Arc ID & RA (J2000)   & Dec (J2000) & Slit angle  &
Slit widths & UVB/ VIS/ NIR exposures  & Date \\ 
        &         &              &             & (deg)       &
(\arcsec) & (s) & \\
\hline

Abell 1689 & ~4.1      & 13:11:32.17 & --01:20:57.3 &
\multirow{2}{*}{--19.4} & \multirow{2}{*}{ $1.3/ 1.2/ 1.2$} &
\multirow{2}{*}{2$\times$3000 / 2$\times$3000 / 6$\times$520} & \multirow{2}{*}{22--03--2010} \\
\smallskip

Abell 1689 &  ~8.1      & 13:11:32.30 & --01:20:50.8\\
Abell 1689 & ~9.1      & 13:11:30.30 & --01:19:48.33  & \multirow{2}{*}{~~18.8} & \multirow{2}{*}{ $1.3/ 1.2/ 1.2$} &\multirow{2}{*}{2$\times$3000 /2$\times$3000 / 6$\times$520} & \multirow{2}{*}{22--03--2010} \\
\smallskip
Abell 1689 & 31.1     & 13:11:30.42 & --01:19:51.5\\

SMACS\,J0304.3--4402 & 1.1 &03:04:20.29 & --44:02:27.8 &~~~~90 & $1.0/ 0.9/ 0.9$ &
4$\times$900 /4$\times$900 / 4$\times$900 & 30--08--2011 \\
MACS\,J0329.6--0211 & 1.2 &03:29:40.18 & --02:11:45.8 &~~~112 & $1.0/ 0.9/ 0.9$ &
4$\times$1200 /4$\times$1200 / 4$\times$1200 & 30--08--2011 \\
SMACS\,J0359.2--7205 & 1.1   &03:59:13.46  & --72:05:14.6  &~~~~49 & $1.0/ 0.9/ 0.9$ &
3$\times$900 /3$\times$900 / 3$\times$900 & 30--08--2011 \\

MACS\,J0416.1--2403 & 1.1 & 04:16:09.85 & --24:03:42.5 & --160 & $1.0/ 0.9/ 0.9$ &
8$\times$800 /8$\times$770 / 8$\times$900 & 13--01--2011 \\
MACS\,J1115.8+0129  & 1.1  & 11:15:51.40 &  01:29:36.6   & ~~--30 & $1.0/ 0.9/ 0.9$ &
4$\times$800 /4$\times$770 / 4$\times$900 & 12--01--2011 \\
MACS\,J1206.2--0847 & 2.1/3.1 & 12:06:14.47 & --08:48:33.3 & ~~~~40 & $1.3/ 1.2/ 1.2$ &
4$\times$1070 / 4$\times$1100 / 4$\times$1200 & 03--04--2011 \\
MACS\,J1311.0--0310 & 1.1 & 13:11:01.37 & --03:10:50.3 & ~~~~90 & $1.0/ 0.9/ 0.9$ & 
4$\times$800 / 4$\times$770 / 4$\times$900 & 03--04--2011 \\
SMACS\,J2031.8--4036 & 1.1  & 20:31:52.89 & --40:37:32.6 & ~~~~~~0& $1.0/ 0.9/ 0.9$ &
 4$\times$800 / 4$\times$770 / 4$\times$900 & 11--10--2010 \\
\smallskip
MACS\,J2129.4--0741& ~$1.5$ & 21:29:26.07 & --07:41:41.6& ~~~~~~0&$1.0/ 0.9/ 0.9$ &
4$\times$800 / 4$\times$770 / 4$\times$900 & 11--10--2010 \\
\hline
\end{tabular}
\label{tab:log}
\end{table*}

\subsection{Further data processing}
The data from the pipeline needed a few more processing steps before a
full analysis could be done. To determine spatial offsets along the
slit between each exposure, one could rely on FITS header information,
or measure offsets directly from the data. We chose the latter, but
because the continuum emission was very faint, this was only possible
after binning the 2D data as described in section
\ref{sect:optbin}. Individually reduced two-dimensional spectra were
combined using their errors as weights.

One-dimensional spectra were extracted by co-adding rows of data. For
each object, the apertures were chosen visually from the binned 2D
spectra with an aperture size matching for the three arms. Typically,
an aperture size of 1--2\arcsec\ was used. Source 8.1 in Abell 1689,
where the X-shooter slit was oriented along the arc, was noticeably
more extended in the 2D spectrum, and an aperture of 4\arcsec\ was
used.

Flux calibration was performed with standard procedures in
IRAF. Observations of various spectrophotometric standard stars
(GD\,71, LTT\,3218, and EG\,274) with a 5\arcsec\ wide slit taken at
the night of the observations were used to calibrate the transmission
in each of the arms. Reference fluxes for these stars covering the UV
to the near-IR were taken from \citet{bohlin01} and
\citet{vernet08}. A correction for an average atmospheric extinction
measured at Paranal \citep{patat11} was applied to the data. M1206,
M1311, M2031, and M2129 were observed in non-photometric conditions.
Since the fluxes in the X-shooter spectrum were corrected to absolute
values by the magnitudes measured in \emph{HST} images, flux
calibration errors would be partly mitigated as the line fluxes were
corrected for slit losses. Even though the slit orientation was not
equal to the parallactic angle, differential slit losses should be
minor effect, since X-shooter has an atmospheric dispersion corrector
in the UVB and VIS arms. The near-IR arm does not have a corrector for
atmospheric dispersion, but since the observations were done at
airmasses below 1.6 the effect was also minor.  The X-shooter pipeline
gives as an output the wavelength solution measured in air, which we
correct to vacuum heliocentric frame before deriving
redshifts. Finally, the spectra were corrected for Galactic reddening
using the dust maps of \citet{schlegel98}.

The extracted and flux-calibrated spectra show small offsets in the
continuum flux between the different arms. In particular, the VIS to
the near-IR differ by 10--20\% in some cases, but given the low
signal-to-noise ratios of the spectra, the flux levels were not
adjusted. The continuum emission follows the expectations
over the entire wavelength range as illustrated in
Fig.~\ref{fig:longspec}. In addition to these small offsets, the flux
calibration showed increased errors at the edges of the spectral
ranges in each arm, where the total transmission is very small due to
the dichroic crossover between the arms. The spectra in these affected
regions are excluded in Fig.~\ref{fig:longspec}.

Emission lines were identified by visual inspection, and line fluxes
were measured with {\tt ngaussfit} within IRAF, assuming that the line
profiles have a Gaussian shape. In some cases of very strong lines,
the lines are clearly not Gaussian, but can be fit well with a
combination of two Gaussian functions \citep[see
  also][]{pettini10,christensen10}. In those cases the listed fluxes
correspond to the sum of the two components. Uncertainties of the line
width, height, and continuum level are propagated for the derivation
of emission line flux uncertainties. Some of the faintest lines have a
more uncertain determination of the peak emission wavelength, which
implies an apparent shift in redshift compared to the brightest
lines. The shifts can be up to $\sim$100 km s$^{-1}$, but are
unphysical if we assume that the lines arise in the same medium. We
determine source redshifts from the mean weighted by the line flux,
which overcomes the problem of the small variation of redshift. Balmer
emission lines are affected by underlying stellar absorption. In order
to correct for this, we subtracted a stellar population model
described in Sect.~\ref{sect:SEDfit} from the galaxy spectrum to get a
pure emission line spectrum before fitting the emission lines.

Removal of telluric absorption lines for the low signal-to-noise ratio
data increases significantly the noise in the data. Instead of
correcting the full one-dimensional spectra, which would introduce
additional errors in the spectra, we chose to correct only the
emission line fluxes individually when necessary. To correct for this
absorption, we used data of bright, hot stars, observed immediately
before or after the science exposure and at the same airmass. The
instrument setup was the same, and the data were reduced in the same
way as the science frames. A one-dimensional spectrum of the star was
extracted, normalised, and the 1D science spectrum divided by this
normalised spectrum. Regions where the atmospheric transmission was
less than 5\% was set to zero, and when fitting emission lines, these
regions were excluded.

\subsubsection{Optimal 2D spectral binning}
\label{sect:optbin}
Observations in the near-IR from the ground are heavily affected by the
strong sky emission lines which vary with time. However, blue-wards of
2.2 micron, the continuum emission from the sky between the sky lines
presents a minor contribution. This is the purpose of the design of
X-shooter; at a sufficiently high spectral resolution, one can
eliminate the regions of sky lines, and co-add the remaining part of
the spectrum to obtain a much deeper detection limit than would
otherwise have been possible as long as the detector read noise does
not dominate over the background sky emission.

With this basic principle, we can use the X-shooter data to detect
very faint objects. In practise, we take the 2D spectra and the
associated 2D error spectra and bin the data by any required amount
using a weighted mean algorithm. This binning effectively smoothes out
any spectral characteristics, so depending on the type of
investigation, we choose a binning factor between 1~{\AA}/pixel (for
investigating absorption lines and very weak emission lines), and
$\sim200$~{\AA}/pixel (for investigating the continuum emission from the
faintest targets).  The latter flux calibrated one-dimensional spectra
for the arcs are presented in Fig.~\ref{fig:longspec}.

\begin{figure*}
\begin{center}
\includegraphics[bb= 80 100 528 740, clip, width=16cm]{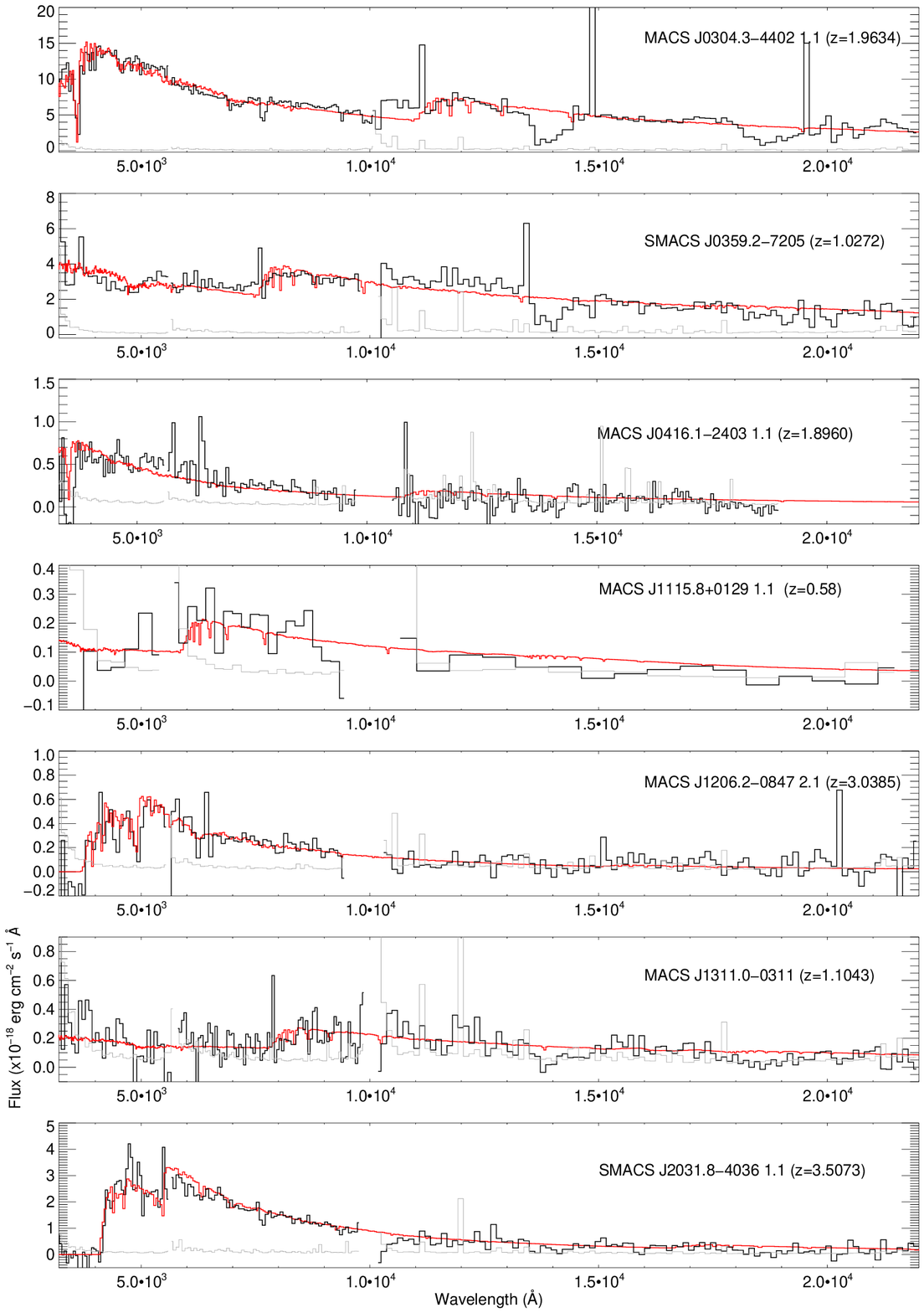}
\end{center}
\caption{Binned source spectra with the best fit spectral templates
  overplotted by the red lines (Sect.~\ref{sect:SEDfit}). The grey
  lines are the error spectra.  Regions with very noisy data are
  omitted, both in the fitting and in the display. In particular,
  wavelengths at 1$\mu$m between the VIS and the near-IR arm are
  dominated by uncertainties in the flux calibration.  \textit{See the
    electronic issue of the paper for a colour version of this
    figure.}}
\label{fig:longspec}
\end{figure*}

\begin{figure*}
\begin{center}
\addtocounter{figure}{-1} 
\includegraphics[bb= 80 275 528 740, clip, width=16cm]{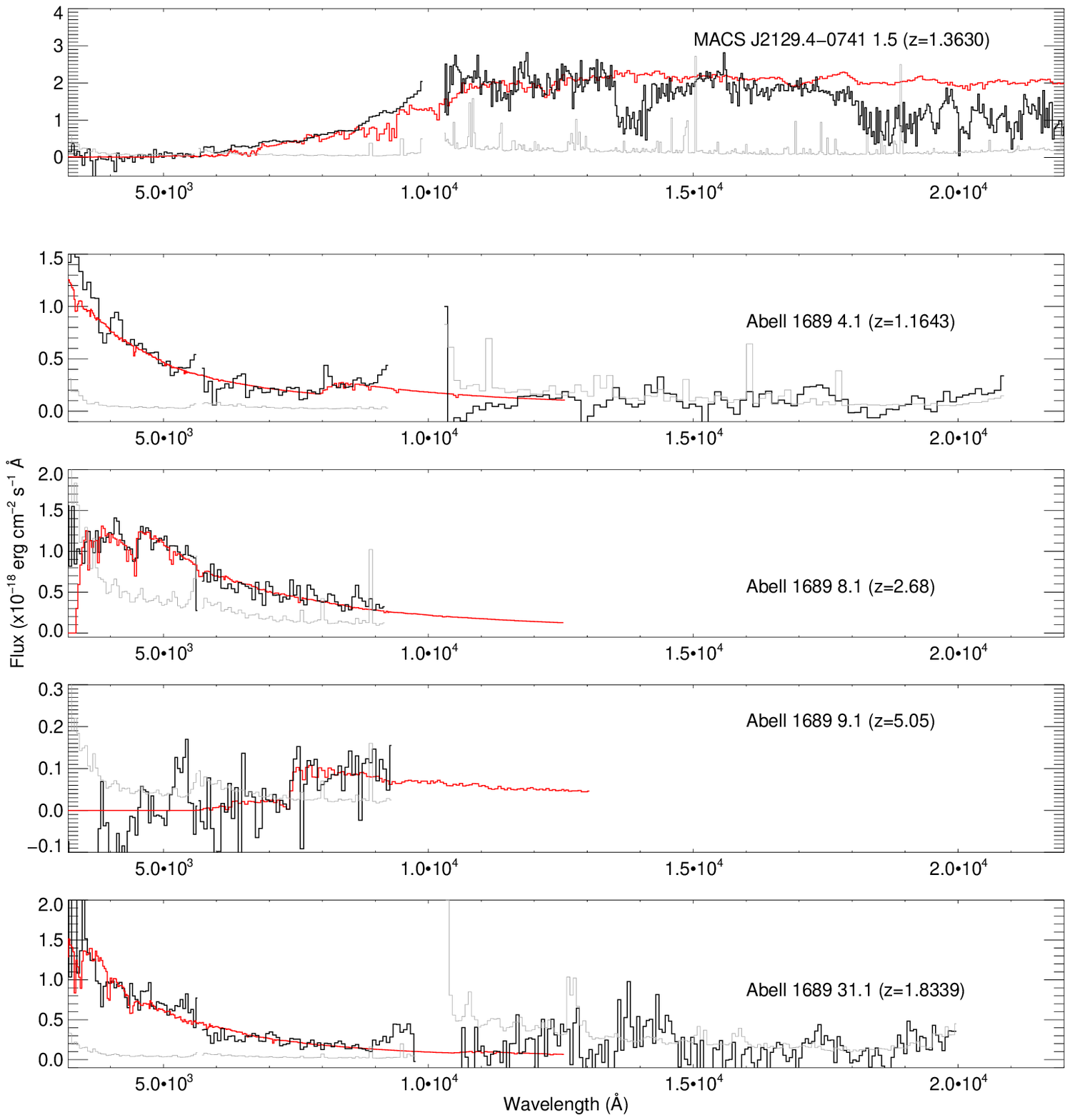}
\end{center}
\caption{ -- \textit{continued.}  The continuum emission in the
    near-IR for the Abell 1689 arcs, is considerably more noisy
    compared to the other spectra due to increased sky subtraction
    errors.}
\end{figure*}

\subsection{High resolution images}
Since the targets are lensed by well-studied massive galaxy clusters,
most of these have extensive high spatial resolution images already
available in the \emph{HST} archive. Both deep and snapshot images
with the WFPC2, ACS or WFC3 instruments have been taken. We obtained
these images from the \emph{HST} archive, and used SExtractor
\citep{bertin96} to derive the photometry of the sources as listed in
Table~\ref{tab:phot}.  Up to date zeropoints were used, and the fluxes
were corrected to total magnitude measured in the detection band ($I$
band).  For the most extended arcs SExtractor was first used to
subtract the extended emission from brighter galaxies in the field to
create a residual image free of emission from the lens galaxies. Then
we defined a polygon around the extended arc in the deepest image and
derived the magnitude in the other filters using the same polygon as
reference.  The photometric uncertainties were determined from from
photon noise measured in the original images before drizzling. In the
case of Abell 1689 ID 9.1, the photometry was defined for a point
source as the 3$\sigma$ limiting magnitude in the measured aperture.

\begin{table*}
\caption{Total AB magnitudes of the arcs.}
\begin{tabular}{lllllllll}
\hline
\hline
Cluster & Arc ID  & F475W & F606W & F625W   & F775W & F814W & F850LP & Camera\\
\hline
A1689   & 4.1      & 24.29$\pm$0.11 & -- & 24.06$\pm$0.10 &
23.78$\pm$0.08 & -- &  23.42$\pm$0.13 & ACS\\
A1689   & 8.1      & 23.29$\pm$0.06 & -- & 22.69$\pm$0.04 &
22.45$\pm$0.04 & -- & 22.40$\pm$0.08 & ACS\\
A1689   & 9.1      & $>$28.2        & -- & $>$27.9        &
25.75$\pm$0.09 &-- & 25.89$\pm$0.15 & ACS\\
A1689   & 31.1     & 24.02$\pm$0.07 & -- & 24.27$\pm$0.12 &
24.08$\pm$0.11 &-- & 24.74$\pm$0.29 & ACS\\
M0304   & 1.1  &-- & 20.19$\pm$0.06 &    &                 &   &                & ACS \\
M0359   & 1.1      &    & 21.58$\pm$0.05 & -- & -- & -- & -- & WFPC2    \\
M0416   & 1.1      & -- & 23.27$\pm$0.09 &  -- & -- & 23.73$\pm$0.12 & --
&   WFPC2\\
M1115 & 1.1 &  23.00$\pm$0.41  & 23.80$\pm$0.10 & 22.94$\pm$0.37 &
23.69$\pm$0.40 & 23.40$\pm$0.6 & --  &  ACS  \\
M1206 & 2.1 &  24.91$\pm$0.05 & 23.16$\pm$0.05 & 24.24$\pm$0.05 & 23.12$\pm$0.02& 22.99$\pm$0.01 & 23.19$\pm$0.03 & ACS\\
M1311 & 1.1 & & 22.58$\pm$0.13 & & & & & WFPC2\\
M2031 & 1.1 & -- & 21.19$\pm$0.10 & -- & -- & -- & -- & ACS\\
M2129 & 1.5 &    -- & -- & -- &  23.48$\pm$0.14 & 23.04$\pm$0.09 & 22.69$\pm$0.15 & ACS\\
\hline
\hline
& & F105W & F110W & F125W & F140W & F160W & & Camera \\
\hline
M0329     & 1.1/1.2    &         & 25.0$\pm$0.3 & & & & & WFC3 \\
M1115     & 1.1 & --  & 22.94$\pm$0.33 & 22.86$\pm$0.20 & 22.54$\pm$0.18 &
22.20$\pm$0.22 &  & WFC3 \\ 
M1206     & 2.1 & 23.13$\pm$0.02 & 23.13$\pm$0.02 & 23.21$\pm$0.02 & 23.12$\pm$0.02 & 23.10$\pm$0.02 & & WFC3\\
M2129     & 1.5 & 21.50$\pm$0.05 & 21.05$\pm$0.05 & 20.72$\pm$0.05 & 20.33$\pm$0.05 & 19.97$\pm$0.01 & & WFC3\\
\hline
\end{tabular}
\label{tab:phot}
\end{table*}

\subsubsection{Slit losses}
Besides the loss of flux in slit spectra since some emission fall
outside the slit due to the seeing effects and inaccuracy of the
centering on the slit, the lensed images also experience slit losses
due to their extended and arc-like nature. Some of the observed images
appear as compact sources even though they are stretched by the
lensing effects, while other images are very extended low surface
brightness arcs. Therefore we determine in each case the fraction of
light falling within the slit relative to the total flux in each
band. The flux-calibrated spectrum is multiplied with the respective
filter transmission function to derive the magnitude and the slit
losses are calculated for each filter.  The final,
wavelength-independent correction factor to be applied to the spectra,
is the weighted average of the factors in all measured bands,
$f_{\mathrm{slit}}$, as listed in Table~\ref{tab:slitloss}, where the
weight in each band determined from the sum of the square of the
photometric uncertainties in Table~\ref{tab:phot} and the uncertainty
derived from the associated one-dimensional error spectrum.

Properties derived from spectra obtained in non-photometric conditions
are partly recovered from the correction of slit losses, since the
method we adopt does not discriminate between losses of light falling
outside the slit or loss of light due to absorption by
clouds. However, since we only apply a constant offset to the flux
calibration, a wavelength-dependent error caused by the
non-photometric conditions could be introduced. Also, galaxies have
colour gradients, but we do not include corrections based on these
gradients in this investigation.

\begin{table}
\caption{Lens image slit loss- and image magnification factors, $\mu$}
\begin{tabular}{llrr}
\hline
\hline
Lens image & ID  &   $f_{\mathrm{slit}}$ &  $\mu$\\
\hline 
Abell\,1689 & 4.1   &   2.7$\pm$0.1  &  23.0$\pm$3.4\\
Abell\,1689 & 8.1   &   2.9$\pm$0.7  &  38.9$\pm$4.0\\
Abell\,1689 & 9.1   &   1.0$\pm$0.1  &  12.7$\pm$2.8\\
Abell\,1689 & 31.1  &   1.7$\pm$0.1  &  26.6$\pm$3.1\\
SMACS0304   & 1.1      &   3.0$\pm$0.1  &  42.0$\pm$8.0 \\
SMACS0359   & 1.1      &   3.0$\pm$0.1  &  18.0$\pm$6.0 \\
MACS0416    & 1.1      &   2.6$\pm$0.3  &  4.6$\pm$0.2 \\ 
MACS1115    & 1.1      &   5.0$\pm$1.7  &  16.3$\pm$1.4\\
MACS1206    &2.1    &   5.1$\pm$0.2  &  9.2$\pm$0.6 \\
MACS1311    & 1.1      &   12.7$\pm$5.3 &  28.0$\pm$12.0\\
SMACS2031    &1.1    &   4.2$\pm$0.3  &  15.8$\pm$7.0\\
MACS2129    &1.5    &   2.1$\pm$0.1  &  9.4$\pm$2.0 \\
\hline
\end{tabular}
\label{tab:slitloss}
\end{table}

\subsection{Lens models}

High spatial resolution images are crucial information for the
analysis of gravitational lensing. The relative projected positions of
the lens and source galaxies along with spectroscopic (or photometric)
redshifts are necessary ingredients to determine the total mass
distribution of the lenses. The lens models allow us to reconstruct
the original images of the galaxies in the source plane before
lensing, to derive their radius, which is essential to derive
dynamical masses.

In this paper, we use mass models constructed using the Lenstool
software \citep{jullo07}\footnote{publicly available at
  http://lamwws.oamp.fr/lenstool/} based on the multiple images
observed in the HST images. The detailed Lenstool model of A1689 has
been published by \citet{limousin07} and the models for the MACS
clusters will be presented elsewhere (Richard et al. in
preparation). We summarise here the main ingredients of the models.
Our starting point is the set of multiple systems identified in the
HST images (between 1 and 7 systems) which complete the ones presented
by other groups \citep{ebeling09,zitrin11a,zitrin11b,zitrin12}. The
majority of these systems have been confirmed with a measured
spectroscopic redshift, including the ones obtained from X-shooter and
presented in this paper. We use the positions and redshifts of these
systems to constrain a multi-component mass distribution described as
a parametric model, including both cluster-scale and galaxy-scale
components \citep[see][for more details]{richard10}. The Monte-Carlo
Markov Chain sampler created by Lenstool provides us with a family of
best models fitting the constraints, allowing us to derive for each
parameter its 1 $\sigma$ error bar. The estimate on the magnification
factors and their corresponding errors are provided for each target in
Table~\ref{tab:slitloss}. In the case of SMACS0304, the uncertainty in
the cluster redshift (estimated between 0.3 and 0.5) is taken into
account when deriving the magnification factor of the target 1.1 in
this cluster. With the information of the total flux from the galaxies
including the effect of magnification and slit losses, we can derive
the absolute properties of the galaxies.

\section{Source spectra}
\label{sect:results}
In this section we present notes for each of the observed arcs.
Table~\ref{tab:arcz} summarises the measured redshifts for the arcs
from emission and absorption lines, as well as redshifts derived from
breaks in highly binned spectra, when no clear emission or absorption
lines could be distinguished. Individual tables in this section list
the detected emission lines and their fluxes uncorrected for
magnification or slit losses. Typically, only few absorption lines are
found by visual inspection for some of the galaxies, because the S/N
level in the continuum is low, and the absorption redshift corresponds
to the average of the detected lines.  As a consequence of the low S/N
levels, the uncertainties of the absorption line redshifts are higher
than those based on emission lines.

\begin{table*}
\centering
    \caption{Cluster and arc redshifts}
  \begin{tabular}{llcccc}
\hline
\hline
Cluster &  $z_{\mathrm{cluster}}$  & Arc ID & $z_{\mathrm{em}}$  & $z_{\mathrm{break}}^d$& 
$z_{\mathrm{abs}}$ \\
\hline
A1689                & 0.181 & ~~4.1    & 1.1643$\pm$0.0001$^a$ &   & \\
A1689                &       & ~~8.1    &                       & 2.68$\pm$0.04 & \\
A1689                &       & ~~9.1    &                       & 5.05$\pm$0.02 &        \\
A1689                &       &  31.1    & 1.8339$\pm$0.0003$^b$ & & 1.8348$\pm$0.0012\\
SMACS\,J0304.3--4402 & 0.3--0.5 & ~~1.1   & 1.9634$\pm$0.0002   & & 1.9635$\pm$0.0005\\
MACS\,J0329.6--0211  & 0.45  &  1.1/1.2 &  ---           &  & \\
SMACS\,J0359.2--7205 & 0.296 &  ~~1.1     & 1.0272$\pm$0.0005     & & \\
MACS\,J0416.1--2403 & 0.420 &  ~~1.1      & 1.8960$\pm$0.0003     & &\\
MACS\,J1115.8+0129   & 0.355 & ~~1.1      &                       &
0.58$\pm$0.02 or 3.51$\pm$0.01 &   \\
MACS\,J1206.2--0847  & 0.439 &  2.1/3.1 & 3.0385$\pm$0.0001$^c$ & & 3.0372$\pm$0.0015 \\
MACS\,J1311.0--0310  & 0.494 &  ~~1.1        & 1.1043$\pm$0.00002    & &\\
SMACS\,J2031.8--4036 & 0.331 & ~~1.1    & 3.5073$\pm$0.0002 &    & 3.5061$\pm$0.0013 \\
MACS\,J2129.4--0741  & 0.589 & ~~1.5    & 1.3630$\pm$0.0004  &   & 1.3617$\pm$0.0012 \\ 
\hline
\end{tabular}
\label{tab:arcz}
\begin{flushleft}
\textit{Notes.~} \\ Cluster redshifts are from \citet{abell89} and for
M1115 and M1206 in \citet{ebeling10}, and M2129 in
\citet{ebeling07}. M0359 and M2031 cluster redshifts are based on
ESO/EMMI and FORS archive spectra. The redshift of the M0304 cluster
is measured from photometry.

Other source spectroscopic redshifts: $^a$ \citet{jullo10} report
$z=1.1648$, $^b$ \citet{broadhurst05} report $z=1.83$, $^c$
\citet{zitrin11b} report $z=3.033$, $^d$ Redshifts based on spectral
modelling.\\
\end{flushleft}
\end{table*}

\subsection{Abell 1689 arc ID 4.1}

Two different lensed images (source IDs 4.1 and 8.1) were placed
within the X-shooter slit. The names of the sources are adopted from
\citet{limousin07}. Source ID 4.1 is reported to have $z=1.1648$
\citep{jullo10}, and is located at the lower edge of the slit during
the observations, and sky subtraction errors are high in the near-IR
arm. From the flux-weighted average of the emission lines in
Table~\ref{tab:a4.1}, we measure the redshift $z=1.1643\pm0.0001$.

Among the other strong optical emission lines that fall within the
spectral range, \hbeta\ is behind a strong sky-line, the
[\ion{N}{ii]}\,$\lambda\lambda$6548,6583 and
[\ion{S}{ii]}\,$\lambda\lambda$6717,6730 doublets are affected by
telluric absorption lines so their fluxes are below the detection
limit.

\begin{table}
    \caption{Emission lines from the A1689 arc ID 4.1}
  \begin{tabular}{lcccr}
\hline
\hline
   line       & $\lambda_r$ $^a$ & $\lambda_{\mathrm{obs}}$ & $z$ & flux$^b$ \\
\hline
\ion{[O}{ii]} & 3727.09  & 8066.41      & 1.16427 & 11.8$\pm$0.2\\
\ion{[O}{ii]} & 3729.88  & 8072.44      & 1.16426 & 15.6$\pm$0.2\\
\ion{[Ne}{iii]}& 3869.84 & 8375.75      & 1.16436 &  3.7$\pm$0.5\\
\hdelta       & 4102.92  & 8880.27      & 1.16438 &  1.8$\pm$0.6\\
\hgamma       & 4341.69  & 9396.41$^d$  & 1.16423 &  5.7$\pm$1.1\\
\ion{[O}{iii]}& 4960.29  & 10735.74$^c$ & 1.16434 & 15.4$\pm$1.9\\
\ion{[O}{iii]}& 5008.24  & 10840.01$^c$ & 1.16444 & 39.8$\pm$2.4\\
\halpha       & 6564.63  & 14208.28$^d$ & 1.16437 & 51.5$\pm$4.4\\
\hline        
\end{tabular}
\label{tab:a4.1}

\textit{Notes.~} \\
$^a$ Rest-frame vacuum wavelengths.\\ 
$^b$ Emission line flux in units of $10^{-18}$ \ecs.\\ 
$^c$ Close to sky line.\\  
$^d$ Affected by strong telluric absorption lines.\\
\end{table}

In the 2D spectrum we detect the emission from a third object located
2\arcsec\ north of A4.1. An extended early type galaxy is located at
RA\,=\,13:11:32.14, DEC\,=\,--01:20:55.33 (J2000) in the ACS image,
but it is only partly covered by the slit. After rebinning the 2D
spectrum a clear break is seen at 4400~{\AA}, consistent with a Balmer
jump at the Abell 1689 redshift. In the spectrum of the third object
we detect the \ion{Ca}{ii}\,$\lambda\lambda$3934,3639 absorption
doublet at $z=0.170$ and a single emission line at 7682 {\AA}
consistent with \halpha\ at $z=0.1702$.


\subsection{Abell 1689 arc ID 8.1 (Giant arc)}
Despite being a giant arc, the redshift of image ID 8.1 remains to be
measured accurately. \citet{broadhurst05} give the best fit
photometric redshift $z_{\mathrm{phot}}=2.63\pm0.48$ while
\citet{halkola06} find $z_{\mathrm{phot}}=3.10\pm0.89$ and
\citet{limousin07} determine $z=2.30\pm0.21$ based on their lens
model.

The object is located at the upper edge of the slit, and is affected
by sky subtraction residuals, which limit our ability to get redshift
information from absorption lines. A faint continuum is detected in
the binned 2D spectrum. With a binning factor of 200 and 400 pixels in
the UVB and VIS arms respectively, any weak and narrow emission or
absorption lines are effectively smoothed out. No emission lines are
detected above a 3$\sigma$ level significance in the unbinned
spectrum.

The binned UVB and VIS spectrum of ID 8.1 is extracted within an
4\arcsec\ aperture and displayed in Fig.~\ref{fig:longspec}. The
flux-calibrated spectrum shows no clear Lyman limit break; however we
find indications of a break at 4470 {\AA}. If we in interpret this
break as the onset of the \lya\ forest, it implies a redshift of
$z=2.68\pm0.04$ by fitting stellar population synthesis spectra as
described in Sect.~\ref{sect:SEDfit}. The lack of a detected Lyman
limit break is due to the high noise level at wavelengths shorter than
3500 {\AA}.

\subsection{Abell 1689 arc ID 9.1}
Source ID 9.1 is recognised as an F625W drop-out in the ACS images,
and has a photometric redshift of $5.16\pm0.74$ and an AB magnitude of
25.9 in the F775W band \citep{broadhurst05}. \citet{halkola06} find a
similar photometric redshift 4.97$\pm$0.78, while lens models predict
a lower redshift $z_{\mathrm{model}}=2.69\pm0.27$ \citep{limousin07}.

Two sources (IDs 9.1 and 31.1) were located within the X-shooter slit
as illustrated in Fig.~\ref{fig:piccy}.  The spatial offset between
source IDs 9.1 and 31.1 projected on the sky is 2\farcs5 as measured
in an ACS image.  The A9.1 continuum spectrum is faint, and
corresponds to an AB magnitude of $26.6\pm0.4$ at $\sim$8000~{\AA},
but is detected in the binned frames illustrated in
Fig.~\ref{fig:longspec}, where we identify a break at 7381~{\AA}. If
the break corresponds to the onset of the \lya\ forest it implies
$z=5.05\pm0.02$. No \lya\ emission line nor any other significant
spectral features are present at these faint flux levels.

\subsection{Abell 1689 arc ID 31.1}

The arc ID 31.1 is adopted from \citet{limousin07}, who argued that
the three images of source ID 12 in \citet{broadhurst05} are
inconsistent with being the same galaxy. The source 31.1 was located
close to the edge of the slit, where sky subtraction errors were
higher than the average.

A wealth of emission lines are detected as listed in
Table~\ref{tab:a9.1}.  Excluding \lya\ while weighting the other
emission line redshift by their measured fluxes, we find
$z=1.8339\pm0.0003$, consistent with that reported in
\citet{broadhurst05} and \citet{limousin07}.  The \lya\ line profile
is double peaked with a brighter red than blue component and has an
equivalent width of 40 {\AA}.  We detect several uncommon emission
lines for this source. In particular we draw attention to the
\ion{O}{iii}] $\lambda\lambda$1661,1666 and \ion{C}{iv}
  $\lambda\lambda$1548,1550 doublets, which indicate the presence of a
  strong ionising source. These lines were also detected in another
  lensed galaxy \citep{fosbury03,villar-martin04}.  The
  temperature-sensitive [\ion{O}{iii]} $\lambda$4363 line is rarely
  detected in galaxies at $z>1$.  \citet{yuan09} find this line in
  another lensed galaxy at $z=1.7$, also behind the Abell 1689
  cluster.  In a separate paper we present a more detailed analysis of
  the emission from this unusual galaxy \citep{christensen12}.

\begin{table}
    \caption{Emission lines from the A1689 arc ID 31.1}
  \begin{tabular}{lcccr}
\hline
\hline
   line       &  $\lambda_r$ $^a$   & $\lambda_{\mathrm{obs}}$ & $z$ & flux$^b$ \\
\hline
\lya          &   1215.67  & 3445.77 &  1.8345  & 72.7$\pm$0.7\\
\ion{C}{iv}   &   1548.20  & 4386.81 &  1.83349 & 4.3$\pm$0.4\\
\ion{C}{iv}   &   1550.77  & 4394.22 &  1.83357 & 4.7$\pm$0.3\\
\ion{O}{iii]} &   1660.81  & 4705.82 &  1.83345 & 3.2$\pm$0.4\\
\ion{O}{iii]} &   1666.15  & 4720.63 &  1.83326 & 7.4$\pm$0.3\\
\ion{[C}{iii]}&   1906.68  & 5402.29 &  1.83335 & 9.1$\pm$0.4\\
\ion{C}{iii]} &   1908.73  & 5408.19 &  1.83340 & 4.7$\pm$0.4\\
\ion{[O}{ii]} &   3727.09  & 10563.06&  1.83413 & 9.2$\pm$0.8\\
\ion{[O}{ii]} &   3729.88  & 10569.40&  1.83371 &11.7$\pm$0.9\\
\ion{[Ne}{iii]}&  3869.84  & 10965.28&  1.83352 & 6.4$\pm$2.5\\  
\hgamma       &   4341.69  & 12301.90&  1.83344 &17.0$\pm$2.1\\
\ion{[O}{iii]}&   4364.44  & 12366.36&  1.83344 & 6.2$\pm$1.8\\
\hbeta        &   4862.70  & 13777.99$^d$   & 1.83340 &42.7$\pm$4.9\\
\ion{[O}{iii]}&   4960.29  & 14057.55$^d$   & 1.83402 &63.0$\pm$10.6\\
\ion{[O}{iii]}&   5008.24  & 14194.09$^{c,d}$& 1.83415 &211.5$\pm$5.7\\
\hline        
\end{tabular}
\label{tab:a9.1}

\textit{Notes.~}\\ 
$^a$ Rest-frame vacuum wavelengths.\\ $^b$ Emission
line flux in units of $10^{-18}$ \ecs.\\ $^c$ Close to sky line.\\
$^d$ Affected by strong telluric absorption lines.
\end{table}

\subsection{SMACS\,J0304.3--4402 arc ID 1.1}

The morphology of the lensed galaxy is complex, and likely consists of
several components that are in the process of a merger. In each of the
lensed image counterparts in the \emph{HST} image, five individual
regions are visible.  The X-shooter spectrum of the source includes
three of these regions. Due to blending by the seeing, however, only
two regions are spatially distinguishable in the two-dimensional
spectrum. Many strong rest-frame optical emission lines are detected
from this galaxy at $z=1.9634\pm0.0002$ as listed in
Table~\ref{tab:m0304} including again the \ion{O}{iii}]
  $\lambda\lambda$1661,1666 doublet.  The emission lines from the two
  regions have a velocity offsets of $\sim$100 km s$^{-1}$, while the
  UV continuum appears similar in shape.  The emission line fluxes in
  Table~\ref{tab:m0304} are the total flux in the X-shooter slit
  summed for the two regions. Even though the Balmer lines are very
  bright from this source, \lya\ emission is absent, and its (UV)
  spectrum appears similar to the cB58 spectrum \citep{pettini00}.  A
  more detailed investigation is presented in \citet{christensen12}.

\begin{table}
    \caption{Emission lines from the M0304 arc ID 1.1}
  \begin{tabular}{lcccr}
\hline
\hline
   line       &  $\lambda_r$ $^a$   & $\lambda_{\mathrm{obs}}$ & $z$ & flux$^b$ \\
\hline

\ion{O}{iii]} &   1660.81 &  4921.93      & 1.96357 & 7.1$\pm$1.6\\
\ion{O}{iii]} &   1666.15 &  4937.35      & 1.96333 & 11.4$\pm$1.9\\
\ion{[O}{ii]}  &  2470.22 &  7322.39      & 1.96337 & 8.4$\pm$2.1\\
\ion{[O}{ii]}  &  3727.09 &  11044.78     & 1.96331 & 437.8$\pm$3.4\\
\ion{[O}{ii]}  &  3729.88 &  11053.09     & 1.96332 & 587.2$\pm$3.9\\
H9             &  3836.49 &  11370.55$^c$ & 1.96372 & 29.1$\pm$4.2\\      
\ion{[Ne}{iii]}&  3869.84 &  11468.93$^c$ & 1.96360 & 149.9$\pm$3.8\\
\ion{He}{i} + H8& 3890.17 &  11527.64$^c$ & 1.96320 & 95.3$\pm$3.2\\
\ion{[Ne}{iii]}&  3968.53 &  11761.36     & 1.96359 & 64.3$\pm$3.4\\
H7             &  3971.20 &  11767.48     & 1.96313 & 51.8$\pm$1.7\\
\hdelta        &  4102.92 &  12158.43     & 1.96329 & 128.5$\pm$2.7\\
\hgamma        &  4341.69 &  12866.19     & 1.96333 & 205.9$\pm$1.3\\
\hbeta         &  4862.70 &  14410.72$^c$ & 1.96345 & 500.0$\pm$0.9\\
\ion{[O}{iii]} &  4960.29 &  14700.41$^c$ & 1.96355 & 669.5$\pm$1.9\\    
\ion{[O}{iii]} &  5008.24 &  14841.33$^c$ & 1.96331 & 2357$\pm$3.5\\
\ion{He}{i}    &  5877.28 &  17417.44     & 1.96345 & 47.0$\pm$5.2\\
\halpha        &  6564.63 &  19453.85$^c$ & 1.96336 & 1802.1$\pm$0.6\\
\ion{[N}{ii]}  &  6585.42 &  19513.96$^c$ & 1.96313 & 58.1$\pm$2.7\\
\ion{[S}{ii]}  &  6718.29 &  19910.44     & 1.96355 & 156.8$\pm$1.1\\
\ion{[S}{ii]}  &  6732.67 &  19951.76     & 1.96335 & 102.7$\pm$1.7\\

\hline        
\end{tabular}
\label{tab:m0304}

\textit{Notes.~}\\ 
$^a$ Rest-frame vacuum wavelengths.\\
$^b$ Emission line flux in units of $10^{-18}$ \ecs.\\
$^c$ Affected by telluric absorption lines.\\
\end{table}

\subsection{MACS\,J0329.6--0211 arc ID 1.1/1.2}

The two mirrored lensed images with IDs 1.1. and 1.2, which have a
photometric redshift of $z=6.18^{+0.07}_{-0.05}$ \citep{zitrin12},
were targeted by the observations, which were carried out by both
nodding and offsetting along the slit. The galaxy has a high surface
brightness region at the edges pointing towards each other, and a more
elongated lower surface brightness region in the extended arcs. The
slit orientation covered only the high surface brightness regions,
where the brightest image ID 1.2 has a magnitude of 25.0 AB mag in the
F110W image \citep[see][for photometric data of all lensed
  images]{zitrin12}.

We detect neither emission lines nor the continuum emission even in a
highly binned 2D spectrum. With the faintness of the galaxy, we do not
expect to detect the continuum emission even with a very large binning
factor, but aim instead for a detection of a single emission line.  We
would be able to detect \lya\ in the VIS arm to a 3$\sigma$
significance level of $5\times10^{-18}$ \ecs\, assuming a typical line
width of 50 km~s$^{-1}$. We cannot exclude that an emission line
happens to fall directly at the wavelength of a strong sky emission
line given the uncertainty of the photometric redshift. In that case,
the 3$\sigma$ detection limit could be much higher: $\sim
2\times10^{-16}$ \ecs.

\subsection{SMACS\,J0359.2--7205 arc ID 1.1}
Emission lines from this arc are listed in Table~\ref{tab:m0359}. No
unusual strong UV emission lines are detected even though its low
redshift ($z=1.0272\pm0.0005$) and the wavelength coverage would allow
for detection of the [\ion{C}{iii}],\ion{C}{iii}]
  doublet. [\ion{O}{iii}]\,$\lambda$4959 lies in a very noisy region
  where the transmission of the NIR arm is low and is not detected.

\begin{table}
    \caption{Emission lines from the M0359 arc ID 1.1}
  \begin{tabular}{lcccr}
\hline
\hline
   line       &  $\lambda_r$$^a$   & $\lambda_{\mathrm{obs}}$ & $z$ &
   flux$^b$ \\
\hline
\ion{[O}{ii]} &   3727.09  & 7558.09      & 1.02788 &  97.7$\pm$3.2\\
\ion{[O}{ii]} &   3729.88  & 7560.08      & 1.02690 & 137.9$\pm$3.4 \\
\ion{[Ne}{iii]}&  3968.53 &  8048.88      & 1.02818 &   5.0$\pm$1.7\\
\hdelta       &   4101.74  & 8317.41      & 1.02778 &  11.0$\pm$2.5\\
\hgamma       &   4341.69  & 8800.95      & 1.02708 &  37.1$\pm$1.6\\
\hbeta        &   4862.70  & 9856.80      & 1.02702 &  95.4$\pm$1.4\\
\ion{[O}{iii]}&   5008.24  & 10152.15     & 1.02709 &  87.9$\pm$4.0 \\
\ion{[N}{ii]} &   6549.91  & 13276.12$^c$ & 1.02692 &  23.8$\pm$3.8\\
\halpha       &   6564.63  & 13306.85$^c$ & 1.02705 & 369.1$\pm$0.9\\
\ion{[N}{ii]} &   6585.42  & 13348.02$^c$ & 1.02690 &  55.4$\pm$4.4\\
\ion{[S}{iii]}&   9533.20  & 19324.30$^c$ & 1.02705 &  56.4$\pm$4.0\\
\hline        
\end{tabular}
\label{tab:m0359}

\textit{Notes.~} \\
$^a$ Rest-frame vacuum wavelengths.\\
$^b$ Emission line flux in units of $10^{-18}$ \ecs.\\
$^c$ Affected by telluric absorption lines.\\
\end{table}

\subsection{MACS\,J0416.1--2403 arc ID 1.1}

Emission lines detected from this arc at $z=1.48960\pm0.0003$ are
listed in Table~\ref{tab:m0416}. \hbeta\ and the [\ion{O}{iii}]
$\lambda\lambda$4959,5007 doublet are strongly affected by telluric
absorption lines, and the correction for the atmospheric absorption
implies a correction factor between 3 and 10.  The flux uncertainty
only reflects the emission line fitting, since the error for the
correction for telluric lines is not propagated. Again, no strong UV
emission lines are detected, because even the strongest rest-frame
optical lines are relatively faint.

\begin{table}
    \caption{Emission lines from the M0416 arc ID 1.1}
  \begin{tabular}{lcccr}
\hline
\hline
   line       &  $\lambda_r$$^a$   & $\lambda_{\mathrm{obs}}$ & $z$ &
   flux$^b$ \\
\hline
\ion{[O}{ii]} &   3727.09  & 10793.11 & 1.89585& 17.2$\pm$1.4\\
\ion{[O}{ii]} &   3729.88  & 10801.18 & 1.89585& 24.5$\pm$1.4\\
\hbeta        &   4862.70  & 14081.70$^c$ & 1.89586& 27.4$\pm$2.1\\
\ion{[O}{iii]}&   4960.29  & 14367.79$^c$ & 1.89656& 27.2$\pm$2.8\\             
\ion{[O}{iii]}&   5008.24  & 14503.75$^c$ & 1.89598& 100.2$\pm$4.2\\
\hline        
\end{tabular}
\label{tab:m0416}

\textit{Notes.~} \\
$^a$ Rest-frame vacuum wavelengths.\\
$^b$ Emission line flux in units of $10^{-18}$ \ecs.\\
$^c$ Affected by strong telluric absorption lines.\\
\end{table}

\subsection{MACS\,J1115.8+0129 arc ID 1.1}
The image of the source is a very low surface brightness arc.  The
X-shooter slit was oriented across the arc so the slit losses were
substantial.

We do not detect any emission lines in either of the arms. The
continuum emission is the faintest among our sample, and is only
recognisable after binning the spectrum by a factor $>$20 along the
dispersion. The binned extracted spectrum in Fig.~\ref{fig:longspec}
is extracted with an aperture of 1\farcs5.  The NIR spectrum is also
very faint, but the continuum emission is detected in the binned
spectrum although it is too faint to be detected in the $K$
band. Between the UVB and VIS spectra, there appears to be a break,
which we can interpret either as the Balmer decrement at
$z\approx0.5$, or a higher redshift for the onset of the \lya\ forest
at $z\approx3.5$. The exact wavelength of the break is sensitive to
the binning of the data; a smaller binning suggests a higher break
wavelength, and thus a higher redshift. Leaving the redshift as a free
parameter in HyperZ \citep{bolzonella00}, we determine the galaxy
redshift $z=0.58\pm0.02$ or $z=3.51\pm0.01$.  The physical properties
of the galaxy derived in Sect.~\ref{sect:specprop} depend on the
assumed redshift, and we suggest that the lowest redshift is valid one
based on the lensing model.

\subsection{MACS J1206.2--0847 arc IDs 2.1 and 3.1}

The conditions during the observations were poor with changing seeing
and thin cirrus clouds. The slit orientation was chosen such that the
two components, denoted by source IDs 2.1 and 3.1 in
\citet{zitrin11b}, which are separated by 2\farcs3, were placed within
the slit. The two sources have the same redshift \citep{zitrin11b}.

\begin{figure}
\begin{center}
\includegraphics[bb=24 305 600 665, clip, width=8.5cm]{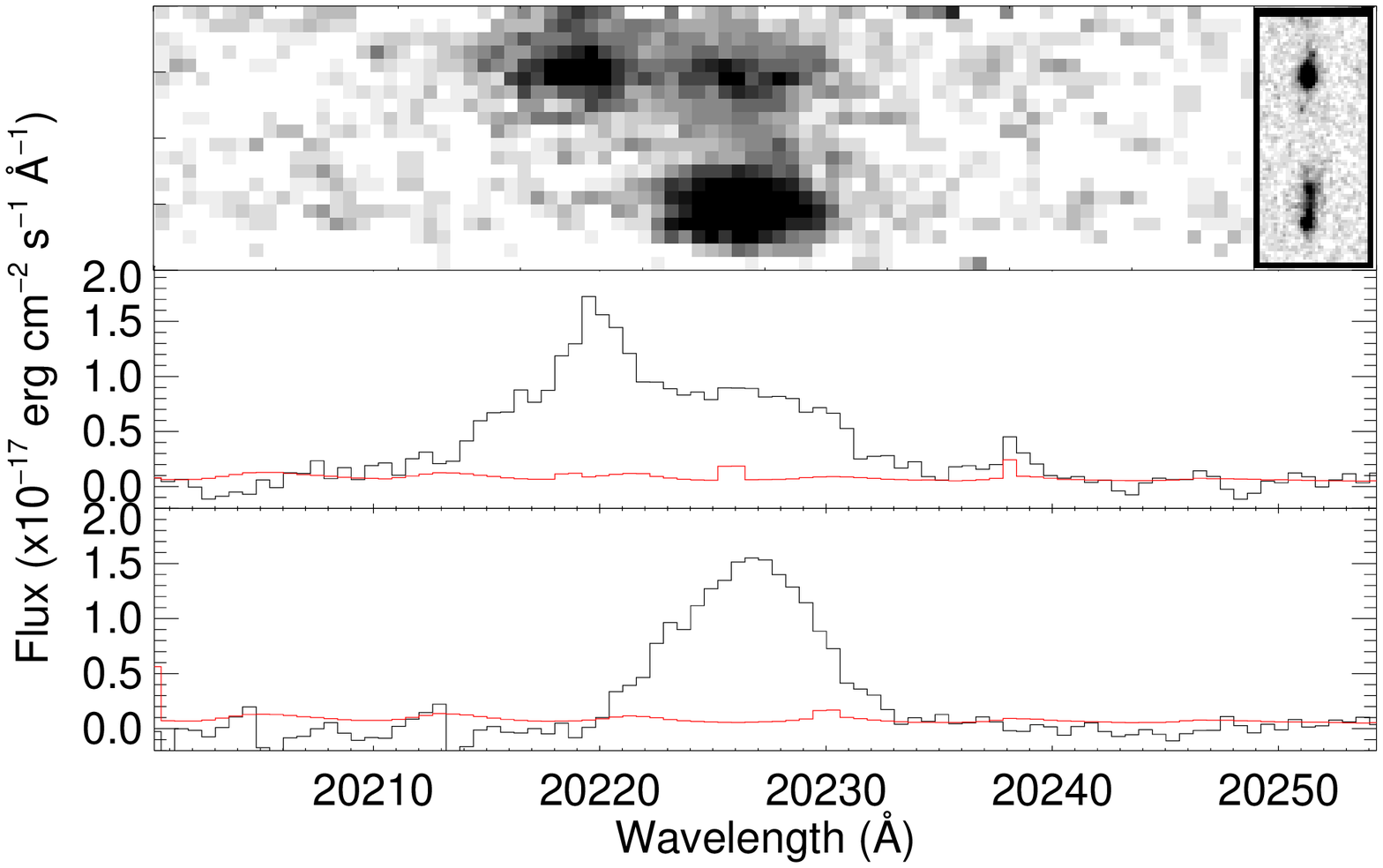}
\end{center}
\caption{\textit{Upper panels}: Two-dimensional X-shooter spectrum
  around the single emission line detected for the M1206 source, and
  the corresponding region around the source in an \emph{HST}/ACS/F606W
  image. The two source IDs 2.1 and 3.1 are separated by 2\farcs3 and
  are seen in both panels. Source 2.1 to the North-East in
  Fig.~\ref{fig:piccy} (upper part in this image) has a compact
  morphology, and source 3.1 to the South-Western one is more
  extended.  The one-dimensional spectra in the middle and lower
  panels are extracted for each of the sources and are corrected for
  telluric absorption lines.}
\label{fig:m1206_cut}
\end{figure}

\begin{figure}
\begin{center}
\includegraphics[bb=25 275 643 677, clip, width=8.5cm]{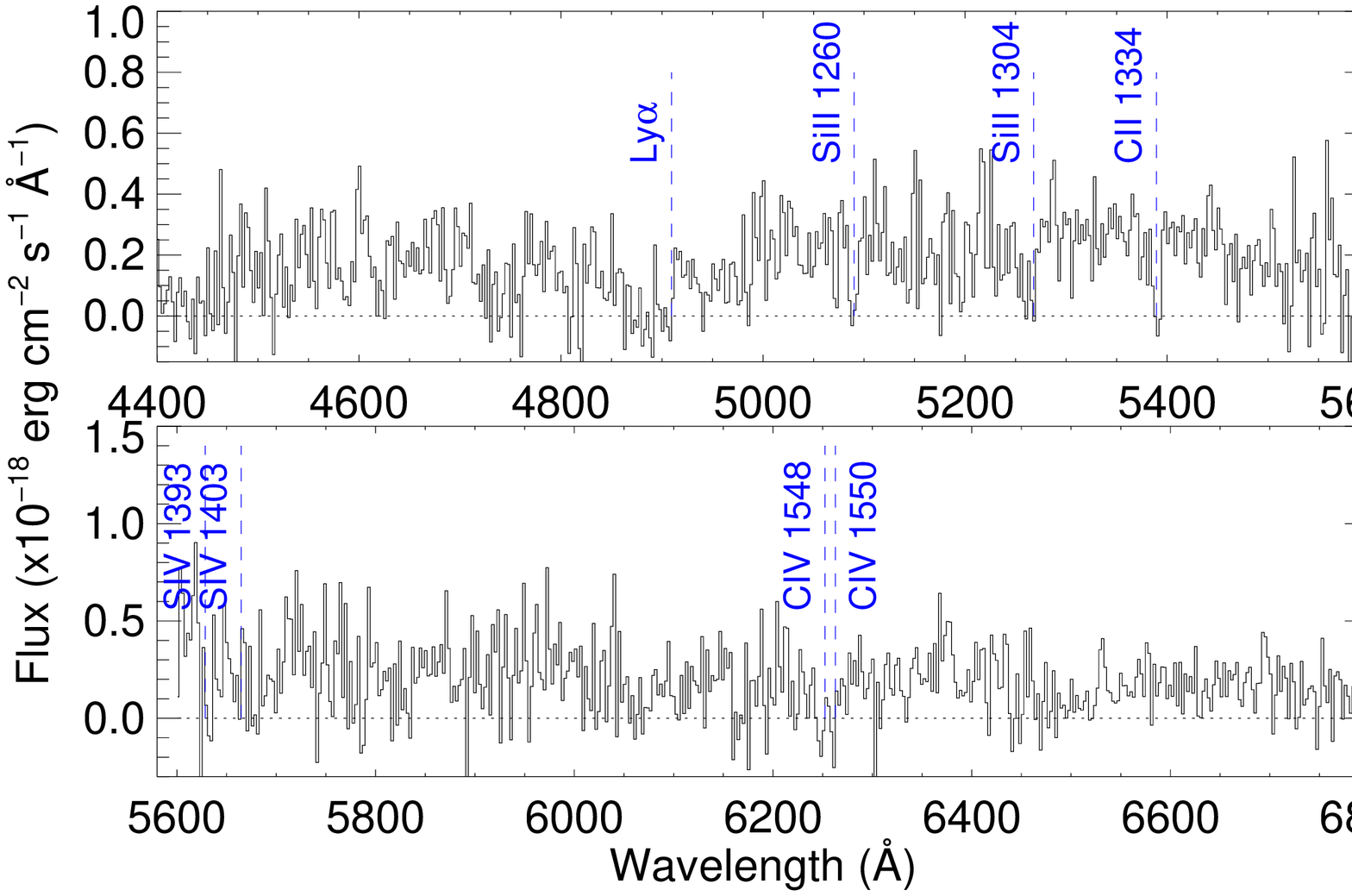}
\end{center}
\caption{Section of the UVB and VIS spectra of the M1206 source ID
  2.1 binned by 10 and 20, respectively, to a dispersion of 2.5~{\AA}
  pixel$^{-1}$. Typical strong UV absorption lines seen in high
  redshift galaxies are marked at the redshift $z=3.0385$.}
\label{fig:m1206_abs}
\end{figure}

We only detect a single, but remarkably bright emission line at
20220~{\AA}, as illustrated in Fig.~\ref{fig:m1206_cut}, and a break
in the continuum around 3700~{\AA}. The line is identified as
\ion{[O}{iii]}\,$\lambda$5007 at $z=3.038$. At this redshift, all the
other strong rest-frame optical emission lines
(\ion{[O}{ii]}\,$\lambda\lambda$3727,3730, \hbeta, and
\ion{[O}{iii]}\,$\lambda$4959) fall at the wavelengths of sky emission
lines or are affected by telluric absorption. At weaker significance,
we detect absorption lines at $z=3.0372\pm 0.0015$ in the binned
spectra illustrated in Fig.~\ref{fig:m1206_abs}.  The marked lines are
the strongest UV lines detected in other high redshift galaxies, such
as in the cB58 galaxy \citep{pettini02}. It appears that \lya\ is seen
in emission at the red wing of a damped \lya\ absorption trough,
however, the emission line is less significant, and cannot be
discerned in the unbinned one- or two-dimensional spectra.  The
inferred redshift is consistent with that measured with low-resolution
optical spectra only \citep{zitrin11b}.

The emission line profile of the compact north-east image ID 2.1
appears to be double-peaked, while the source ID 3.1 only shows one as
illustrated in Fig.~\ref{fig:m1206_cut}. The separation of the two
emission components in the 2.1 source is 7.9$\pm$0.3~{\AA}
corresponding to a velocity separation of 117$\pm$5 km~s$^{-1}$.

The emission line fluxes in Table~\ref{tab:m1206} include the
uncertainties from fitting Gaussian profiles to the observations, but
they do not include the uncertainty due to the non-photometric
conditions during the observations.  The line fluxes are derived after
correction for telluric lines, which absorb between 40 and 60\% of the
emission.

\begin{table}
    \caption{Emission lines from the M1206 arc. Source IDs are adopted
    from \citep{zitrin11b}.}
  \begin{tabular}{lcccr}
\hline
\hline
   line       &  $\lambda_r$ $^a$   & $\lambda_{\mathrm{obs}}$ & $z$ & flux$^b$ \\
\hline
Source ID 2.1: \\
\ion{[O}{iii]}   &   5008.24  & 20217.87 & 3.03696 & 83.4$\pm$1.4\\
\ion{[O}{iii]}   &   5008.24  & 20225.77 & 3.03854 & 101.9$\pm$1.8\\
\hline
Source ID 3.1:
\ion{[O}{iii]}   &   5008.24  & 20226.02  & 3.03859 & 119.7$\pm$1.2\\
\hline
\end{tabular}
\label{tab:m1206}

\textit{Notes.~}\\ $^a$ Rest-frame vacuum wavelengths.\\  
$^b$ Emission line flux in units of $10^{-18}$ \ecs. 
\end{table}

\subsection{MACS\,J1311.0--0310 arc ID 1.1}

From the emission lines listed in Table~\ref{tab:m1311} we derive a
weighted redshift average of $z=1.10433\pm0.00002$. At this redshift
\halpha\, and [\ion{N}{ii}] fall in the region between the $J$ and $H$
bands where telluric absorption is severe. The observing conditions
were non-photometric, which affect the absolute values of the emission
lines. We note that the ratios of emission lines is most likely not
affected, because the transmission derived from the spectrophotometric
standard star observed during the night appeared similar in shape to
that of other nights.

\begin{table}
    \caption{Emission lines from the M1311 arc}
  \begin{tabular}{lcccr}
\hline
\hline
   line       &  $\lambda_r$ $^a$   & $\lambda_{\mathrm{obs}}$ & $z$ & flux$^b$ \\
\hline
\ion{[O}{ii]}   & 3727.09  & 7843.14  & 1.10436 & 8.0$\pm$1.3\\ 
\ion{[O}{ii]}   & 3729.88  & 7848.85  & 1.10432 & 12.0$\pm$0.4\\
\hbeta        &   4862.70  & 10232.73 & 1.10433 & 30.6$\pm$2.6\\
\ion{[O}{iii]}&   5008.24  & 10539.05 & 1.10434 & 13.9$\pm$2.4\\
\hline
\end{tabular}
\label{tab:m1311}

\textit{Notes.~}\\ $^a$ Rest-frame vacuum wavelengths.\\
$^b$ Emission line flux in units of $10^{-18}$ \ecs.
\end{table}

\subsection{SMACS\,J2031.8--4036 arc ID 1.1}

Several emission lines are detected from this arc as listed in
Table~\ref{tab:m2031}, and the flux-weighted average redshift is
$z=3.5073\pm0.0002$. Most lines are not affected by telluric
absorption, but \ion{H}{i}\,$\lambda$3971 has a significant
correction, and the flux in Table~\ref{tab:m2031} gives the flux after
the correction is applied. Apart from that, telluric absorption and
sky emission lines affect our ability to detect \hgamma\ and \hdelta,
which fall in the region between the $H$ and $K$ bands.

Again in this case, the \lya\ emission line profile is double-peaked,
and a wealth of information can be derived from the UV emission lines,
which is presented in a separate paper \citep{christensen12}.  The
redshift listed in Table~\ref{tab:m2031} for \lya\ reflects the
redshift of the bluest wavelength of the red peak as identified by
visual inspection.

At the low S/N per pixel in the UVB and VIS arms, we can only identify
several Lyman forest absorption lines, and \ion{Si}{ii}\,$\lambda$1206
and \ion{S}{iv} $\lambda\lambda$1394,1403 at $z=3.5$. A detailed
analysis of the absorption lines requires a spectrum with a higher S/N
measured in the continuum.

\begin{table}
    \caption{Emission lines from the M2031 arc}
  \begin{tabular}{lcccr}
\hline
\hline
   line       &  $\lambda_r$ $^a$   & $\lambda_{\mathrm{obs}}$ & $z$ & flux$^b$ \\
\hline
\lya           &   1215.67  & 5480.80 &  3.50845 &173.7$\pm$0.5 \\
\ion{N}{iv]}   &   1486.50  & 6700.85 &  3.50780 & 7.6$\pm$0.9 \\
\ion{O}{iii]}  &   1660.81  & 7485.15 &  3.50693 & 2.9$\pm$0.6 \\
\ion{O}{iii]}  &   1666.15  & 7509.46 &  3.50707 & 8.8$\pm$0.7 \\
\ion{[C}{iii]} &   1906.68  & 8594.04 &  3.50733 &12.0$\pm$0.5 \\
\ion{C}{iii]}  &   1908.73  & 8603.54 &  3.50747 & 8.4$\pm$0.9 \\
\ion{[O}{ii]}  &   3727.09  & 16800.18&  3.30759 &14.4$\pm$1.6 \\
\ion{[O}{ii]}  &   3729.88  & 16811.01&  3.50712 &15.5$\pm$1.3 \\
\ion{[Ne}{iii]}&   3869.84  & 17442.96&  3.50741 &14.8$\pm$2.1  \\
\ion{H}{i} (H7)&   3971.20  & 17898.76&  3.50714 & 6.0$\pm$1.0\\ 
\hbeta         &   4862.70  & 21917.47&  3.50726 &37.7$\pm$1.3 \\ 
\ion{[O}{iii]} &   4960.29  & 22357.68&  3.50733 &61.9$\pm$0.9 \\             
\ion{[O}{iii]} &   5008.24  & 22573.68&  3.50731 &205.2$\pm$0.5 \\
\hline        
\end{tabular}
\label{tab:m2031}

\textit{Notes.~} \\
$^a$ Rest-frame vacuum wavelengths.\\
$^b$ Emission line flux in units of $10^{-18}$ \ecs.\\
\end{table}

\subsection{MACS J2129--0741 arc ID 1.5}

The galaxy cluster MACS J2129.4--0741 at $z=0.589$ shows a galaxy
lensed in six distinct images. \citet{zitrin11a} predict a source
redshift of $z=1.0-1.5$ based on their lens model.  The galaxy is red,
has colours similar to the lens galaxies, and its morphology is
consistent with being an early type galaxy.

\begin{figure}
\begin{center}
\includegraphics[bb=20 290 630 555, clip, width=8.5cm]{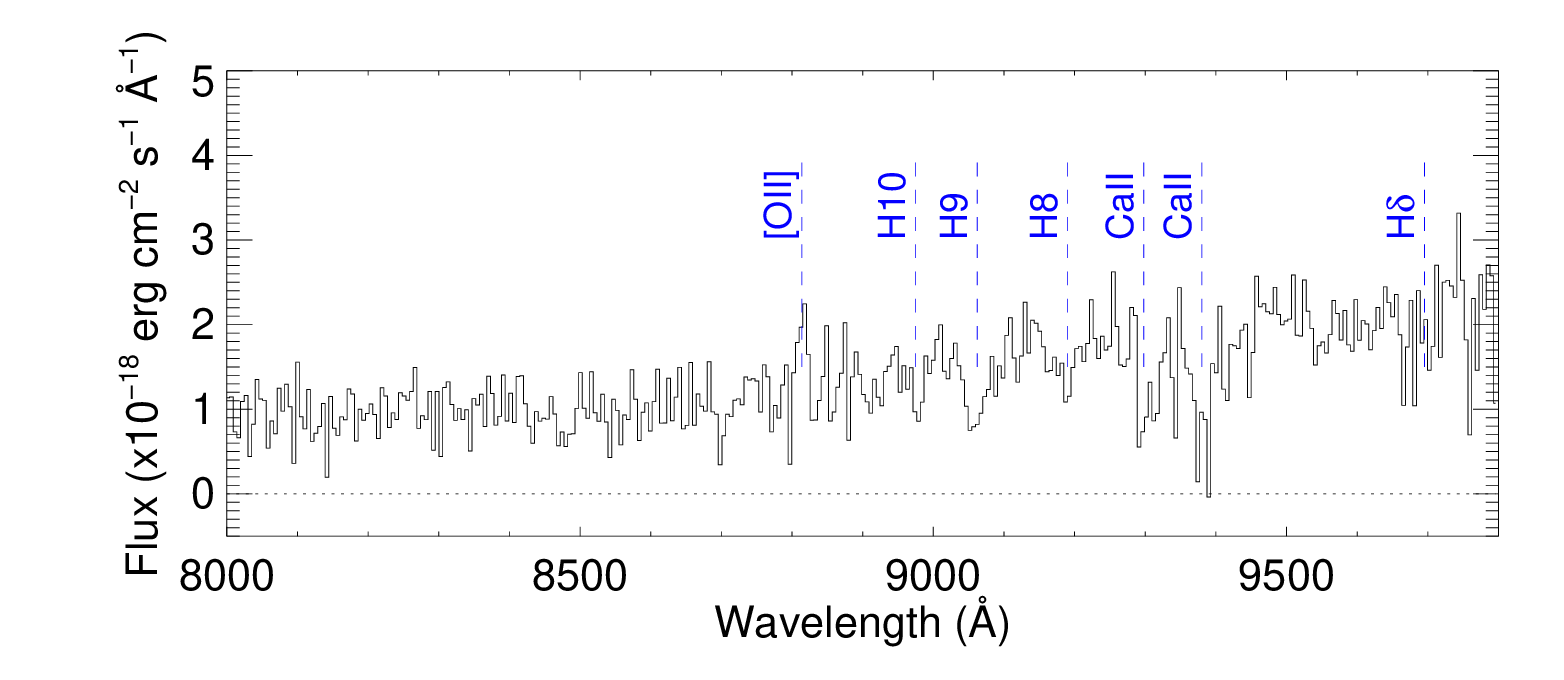}
\end{center}
\caption{Section of the VIS and near-IR spectra of the M2129 arc ID
  1.5 binned to a dispersion of 5.2 and 12~{\AA} pixel$^{-1}$,
  respectively. Detections of \ion{Ca}{ii} 3934,3969 lines and
  absorption lines from the Balmer series of hydrogen at $z=1.3617$
  are indicated. Also a faint emission line from the [\ion{O}{ii}]
  $\lambda3727$ doublet is visible in this binned version.}
\label{fig:m2129_abs}
\end{figure}

For the X-shooter observations, we targeted the isolated,
southern-most image of the six. Although it is not the brightest of
the images, source blending and background subtraction is not a
problem at this location. In the extracted spectra we find weak oxygen
emission lines listed in Table~\ref{tab:m2129}. These lines imply a
redshift $z=1.3630\pm0.0004$. At this redshift, absorption lines from
\ion{Ca}{ii}\,$\lambda\lambda$3934,~3969 are detected after telluric
absorption lines between 9300 and 9500 {\AA} are corrected for, and
several Balmer series absorption lines are also detected (see
Fig.~\ref{fig:m2129_abs}).

\begin{table}
    \caption{Emission lines from the M2129 arc}
  \begin{tabular}{lcccr}
\hline
\hline
   line       &  $\lambda_r$ $^a$   & $\lambda_{\mathrm{obs}}$ & $z$ & flux$^b$ \\
\hline
\ion{[O}{ii]}  &   3727.09  & 8805.93&  1.36251 & 8.6$\pm$1.2 \\
\ion{[O}{ii]}  &   3729.88  & 8814.78&  1.36315 & 7.5$\pm$1.1\\
\ion{[O}{iii]} &   5008.24 & 11834.63&  1.36314 & 36.5$\pm$1.7\\
\hline        
\end{tabular}
\label{tab:m2129}

\textit{Notes.~} \\
$^a$ Rest-frame vacuum wavelengths.\\
$^b$ Emission line flux in units of $10^{-18}$ \ecs.\\
\end{table}

\section{Physical properties from the continuum emission}
\label{sect:specprop}
In this section we derive physical properties based on the continuum
spectra from the sources.

\subsection{Stellar population fit}
\label{sect:SEDfit}

To determine the stellar masses of the galaxies, along with other
parameters, we use spectral energy distribution fitting (SED)
methods. Conventionally, SED fits make use of sparsely sampled
photometric data from broad or narrow band imaging data, whereas we
here are dealing with the entire spectral range from the observed UV
to near-IR. Since the spectra have low S/N ratio, we first bin the
data by a factor between 100 and 1000, depending on the initial
quality of the spectra, giving resolutions in the range $\sim$20--200.
The binned spectra and the associated error spectra are shown in
Fig.~\ref{fig:longspec}.

We compare the spectra with a large set of model spectra from
\citet{bruzual03}. We extract spectral templates with 0.2, 0.4 and 1.0
solar metallicities, and star formation histories described as an
instantaneous burst or exponentially declining star formation rates
(SFR) with e-folding time scales of 10, 50, 100, 300, 500, and 1000
Myr, and exponentially increasing SFRs with time scales of 10, 50 and
500 Myr. The initial mass function (IMF) is either a Salpeter
\citep{salpeter55}, or a Chabrier IMF \citep{chabrier03} with lower
and upper mass cutoffs of 0.1 and 100 M$_{\odot}$, respectively.  The
best fits are determined by minimising
\[\chi^2 = \sum_{\lambda} ((f_{\mathrm{obs},\lambda} -
f_{\mathrm{model},\lambda})/ \sigma_{\mathrm{obs},\lambda})^2,\] where
the sum is taken over all photometric points, $\lambda$ in the
spectra. In practise, there is no significant distinction between the
quality of the fits between the two IMFs. We therefore choose to only
use the Chabrier IMF to derive the stellar masses for consistency.  If
a Salpeter IMF is selected instead, the stellar mass is a factor $1.7$
higher due to the flattening of the Chabrier IMF below 1\,M$_{\odot}$.

To determine the best fit spectral template, the flux at each (binned)
wavelength is converted to an AB magnitude, and we use a modified
version of HyperZ \citep{bolzonella00}, which allows us to include for
each object a catalogue which contains up to 500 individual
photometric data points. From all the photometric points between
3000--22000 {\AA}, we exclude the wavelengths around 1 micron, where
the flux calibration causes inaccuracies at the wavelengths where the
dichroic splits the light between the VIS and near-IR arms. We also
exclude the regions between the $J$--$H$ and $H$--$K$ bands which are
strongly affected by telluric absorption. HyperZ also allows to
determine the best fit reddening as an additional parameter. Here we
use a starburst galaxy extinction curve \citep{calzetti00} as
input. We do not constrain the reddening with that determined from the
Balmer decrements where available (see Sect.~\ref{sect:ebv_gas}),
because the stars may not experience the same amount of reddening as
the gas. For starburst galaxies, \citet{calzetti01} finds that the
radiation from stars and ionised gas experience different extinctions
and that the stellar reddening is on the average 44\% of that in the
gas phase.  The flux attenuation bluewards of \lya\ is taken into
account using the prescription for the transmission in the
intergalactic medium in \citet{madau95}.

For the galaxies with known spectroscopic redshifts, their redshifts
are fixed in the template fitting, while in two cases (MACS\,1115 and
Abell 1689 ID 8.1), the best fit was also used to derive the
photometric redshift.  We do not constrain the template metallicities
to the ones best matching the nebular abundances derived in
Sect.~\ref{sect:R23}, because these may not be the same, as older
stellar populations, if present, may have lower metallicities
reflecting the lower abundances of the gas at their time of
formation. The output gives the best fit template which allows us to
derive the ages, metallicities, and reddening as listed in
Table~\ref{tab:sedfit}. The best fit spectra are overplotted on
Fig.~\ref{fig:longspec}.  One of the outputs of HyperZ is a scaling
factor $b$ allowing the galaxy stellar mass to be derived: \[M_* = b
(1+z) 4\pi D_L^2/L_{\odot,\mathrm{bol}},\] where
$L_{\odot,\mathrm{bol}}$ is the solar bolometric luminosity
(3.826$\times10^{33}$ erg s$^{-1}$) and $D_L$ is the luminosity
distance, where we assume a flat cosmology with
$\Omega_{\Lambda}=0.72$ and $H_0=73$~km s$^{-1}$~Mpc$^{-1}$.  The
derived stellar mass is magnification dependent and is sensitive to
slit losses. Table \ref{tab:sedfit} lists the corrected stellar masses
including the correction factors in Table~\ref{tab:slitloss}.
Uncertainties for the parameters (mass, age, reddening) are determined
by adding to the initial spectrum the error spectrum multiplied by a
random number normalised to a Gaussian distribution with $\sigma=1$. A
thousand fits are performed for each galaxy spectrum, and the standard
deviation of the fitted values are taken as the uncertainties for the
parameters.

In all cases the best fit templates are obtained with a model of an
instantaneous burst of star formation.  Such models are artificial in
the sense that this would mean that there is no current on-going star
formation, which is not correct because the galaxies show strong
emission lines. Other templates that produce acceptable fits are those
with a small value of $\tau=10$~Myr, while in a few cases, templates
extracted from a constant star formation rate is also a valid
solution. In a few cases, an exponentially increasing SFR also
produces an acceptable fit. With these alternative fits the stellar
masses remain the same as those listed in Table~\ref{tab:sedfit} to
within $\pm$10--20\% uncertainty.

As usually the case for SED fitting, there is a strong degeneracy
between the age of the stellar population and the reddening. For any
given spectrum, a similar good fit can be obtained with a younger
population combined with more reddening of the template.  The galaxy
ages as obtained by the best fits are generally small ($\lesssim100$
Myr), and the reddenings are small as well, such that the allowed
parameter space of reddening-age values is limited. One concern is
that, if the ages are underestimated by the SED fit, the stellar
masses will be as well. It is common in the literature to only include
templates in the SED fits that are older than the typical dynamical
timescale for galaxies, i.e. older than a few 10 Myr
\citep[e.g.][]{wuyts12}. For the lensed galaxies in this study, the
resulting fits when forcing the age to be older than 20 Myr, the fits
deteriorate significantly bluewards of the \lya\ wavelengths.
Similarly, if we fit the spectra with templates of a constant star
formation rate and a lower limit for the age of a few times 10 Myr,
the fits at UV wavelengths deteriorate, while the rest of the spectrum
can be fitted well with an older template. We therefore choose to not
constrain the possible template ages in the fitting. Another issue is
that an underlying older stellar population can add to the total
stellar mass, while not be detected because it does not contribute
significantly to the UV radiation.  Since the $K$-band fluxes of the
galaxies are small, we argue that a significantly more massive older
stellar population is not present. To better constrain the exact
masses of an underlying older stellar populations in the galaxies,
deep rest-frame near-IR images are needed to fit the SED with a larger
wavelength range.

\begin{table*}

\caption{Best spectral fit parameters from Sect.~\ref{sect:SEDfit}. The
  masses,  SFRs, and magnitudes are
  corrected for slit losses and lens magnification factors.}
\begin{tabular}{ll@{}clllll}
\hline
\hline
Arc  & Z/Z$_{\odot}$  & Age  & \ebv$_{\mathrm{stars}}$ & log $M_*$  & 
UV SFR & UV SSFR \\ 
     &              & (Myr)&  (mag)   & (M$_{\odot}$)   &
(M$_{\odot}$ yr$^{-1}$) & (Gyr$^{-1}$) \\
\hline

A4.1  & 0.2   & 4$\pm$3   & 0.10$\pm$0.02 &  6.97$\pm$0.08 &
0.4$\pm$0.1 & 40$\pm$11  \\ 

A8.1  & 0.2   & 4$\pm$3   & 0.07$\pm$0.08 &  7.85$\pm$0.15 &
1.6$\pm$0.7 & 23$\pm$11  \\ 

A9.1  & 1.0   & 90$\pm$50   & 0.00$\pm$0.02 &  9.79$\pm$0.13 &
0.6$\pm$0.2 & 0.09$\pm$0.04 \\ 

A31.1 & 0.2   & 4$\pm$2      & 0.17$\pm$0.04 &  7.70$\pm$0.10 &
1.0$\pm$0.2 & 20$\pm$6 \\ 

M0304 & 0.4   & 57$\pm$6    & 0.17$\pm$0.02 & 10.57$\pm$0.08 & 
11.6$\pm$2.3 & 0.30$\pm$0.02 \\ 

M0359 & 0.4   & 110$\pm$18   & 0.15$\pm$0.03    & 9.62$\pm$0.13  &
1.5$\pm$0.5 & 0.36$\pm$0.05 \\ 

M0416 & 1.0   & 32$\pm$14   & 0.00$\pm$0.08    & 8.70$\pm$0.16  &
0.9$\pm$0.2  & 1.8$\pm$0.8 \\ 

M1115 & 1.0   & 400$\pm$260  & 0.00$\pm$0.06 &  7.83$\pm$0.22 &
0.007$\pm$0.003 & 0.10$\pm$0.06 \\ 

M1206 & 1.0   & 20$\pm$13    & 0.00$\pm$0.03 &  9.23$\pm$0.25 &
3.7$\pm$0.6 & 2.2$\pm$1.7 \\ 

M1311 & 1.0   & 140$\pm$20   & 0.15$\pm$0.09 &  9.14$\pm$0.21
&0.3$\pm$0.2 & 0.2$\pm$0.1 \\ 

M2031 & 0.2   & 4$\pm$3      & 0.00$\pm$0.03 &  9.16$\pm$0.21 &
18.0$\pm$8.2 & 13$\pm$6 \\ 

M2129 & 1.0   & 4700$\pm$1500 & 0.07$\pm$0.06 &  11.39$\pm$0.12
&0.003$\pm$0.001 & $10^{-5}$ \\ 

\hline
\end{tabular}
\label{tab:sedfit}
\end{table*}

\subsection{Absolute magnitudes}
A relevant question to ask given the heterogeneous sample, is what
their magnitudes are relative to a typical galaxy at the same
redshift. To calculate this, we use the best fit model spectrum to
derive the magnitudes at rest-frame 1500~{\AA}, and compare these to
the values of $M^*_{\mathrm{AB}}$, which depends on the redshift
\citep{arnouts05}. We use the best fit templates, rather than the
observed spectra because of the low S/N ratio in the spectra.  The
resulting absolute magnitudes and that compared with
$M^*_{\mathrm{AB}}(z)$ are listed in Table~\ref{tab:mag_r}. The errors
reflect the uncertainties in the magnification factor and slit losses.

For the galaxies with a low UV-flux characteristic for an evolved
stellar population, neither their UV absolute magnitude nor their UV
SFRs derived in Sect. ~\ref{sect:sfr}, are relevant for comparing with
star forming galaxies at a given redshift selected in a uniform way
from large surveys. When the spectra show evidence for older stellar
populations it is more reasonable to compare the relative luminosities
at rest-frame visual wavelengths. Table \ref{tab:mag_r} lists the
absolute magnitudes and luminosities of the galaxies at rest-frame
6500~{\AA} corrected for magnification and slit losses. Again the best
fit template is used to determine the luminosities. For the Abell 1689
targets with poor near-IR spectra, and M2031 at $z=3.5$, the 6500
{\AA} flux is based on the best fit template spectra at wavelengths
redwards of the measured spectrum.  The redshift dependent
characteristic magnitudes ($M^*_{\mathrm{AB}}$) defined by the
luminosity function of galaxies in the rest-frame $r'$ band are
adopted from \citet{gabash06}.

Comparing the relative rest-frame UV and optical luminosities, we see
that most galaxies are UV bright. This is by selection, since the
observations targeted galaxies with both high total brightness and
high surface brightness regions, which at higher redshifts implies a
high UV surface brightness. The only exception in our sample is the
M2129 arc, which was known to be an old evolved system based on its
red colour and its morphology.

\begin{table*}
    \caption{Rest-frame optical absolute magnitudes and relative
      luminosities at 1500 and 6500~{\AA}.}
  \begin{tabular}{lllll}
\hline
\hline
 Arc  &   $M_{\mathrm{AB},1500}$ &
 $L/L^*$($z$,1500~{\AA}) & $M_{\mathrm{AB},6500}$ & $L/L^*$($z$,6500~{\AA}) \\
\hline
A4.1  & --18.39$\pm$0.29 & 0.2$\pm$0.1   & --17.93$\pm$0.29 & 0.013$\pm$0.004 \\
A8.1  & --20.56$\pm$0.42 & 0.9$\pm$0.4   & --19.95$\pm$0.42 & 0.07$\pm$0.03\\
A9.1  & --19.97$\pm$0.41 & 0.5$\pm$0.2   & --23.81$\pm$0.41 & 2.4$\pm$1.0\\
A31.1 & --19.76$\pm$0.19 & 0.4$\pm$0.1   & --17.62$\pm$0.19 & 0.008$\pm$0.001\\
M0304 & --22.46$\pm$0.20 & 5.0$\pm$1.0   & --23.44$\pm$0.20 & 1.7$\pm$0.3  \\
M0359 & --19.82$\pm$0.35 & 0.8$\pm$0.3   & --21.42$\pm$0.35 & 0.3$\pm$0.1 \\
M0416 & --19.65$\pm$0.19 & 0.4$\pm$0.1   & --20.60$\pm$0.19 & 0.1$\pm$0.02  \\
M1115 & --13.65$\pm$0.40 & 0.004$\pm$0.002&--16.43$\pm$0.40 & 0.004$\pm$0.002\\
M1206 & --21.55$\pm$0.16 & 1.7$\pm$0.3   &--21.50$\pm$0.16 & 0.24$\pm$0.04 \\
M1311 & --18.22$\pm$0.72 & 0.2$\pm$0.1   &--19.95$\pm$0.72 & 0.08$\pm$0.06\\
M2031 & --23.39$\pm$0.45 & 9.3$\pm$4.2   &--23.00$\pm$0.45 & 1.0$\pm$0.4  \\
M2129 & --13.28$\pm$0.29 & 0.002$\pm$0.001&--22.07$\pm$0.29 & 0.8$\pm$0.2  \\
\hline        
\end{tabular}
\label{tab:mag_r}
\end{table*}


\section{Physical conditions from emission lines}
\label{sect:emphys}

In this section we present properties of the lensed galaxies, which
have sufficiently bright emission lines for a derivation of physical
quantities. Diverse characteristics of the galaxies are expected,
because the galaxies are not selected with a known redshift, nor
appearance, hence we cannot derive a uniform set of properties for all
sources with the current data set. All the physical properties derived
in this section are summarised in Table~\ref{tab:phys_prop}. When line
ratios are considered, we do not apply any correction for slit losses
and magnification factors, since we assume that the emission lines
come from the exact same region.

\subsection{Kinematics}

\subsubsection{Line widths}
\label{sect:dynmass}
We derive the mean velocity dispersion ($\sigma$) from all the
emission line widths, excluding \lya\ when detected, and weighted with
their associated uncertainties. The measured line widths are corrected
for instrumental resolution by measuring the widths of nearby
unblended sky lines, and subtracting the widths in quadrature. The
results are listed in Table~\ref{tab:phys_prop}. To determine the
dynamical masses of the galaxies, their sizes need to be measured
after lens modeling and source reconstruction, and will be presented
elsewhere (J. Richard in prep.). Although not included in
Table~\ref{tab:phys_prop} because only few lines are detected, we can
still determine the velocity dispersion of the emission lines from
M2129 ID 1.5 to be 147$\pm$17 km s$^{-1}$, and 40$\pm$3 km s$^{-1}$
for M1206 ID 2.1. The small velocity dispersions derived from the
emission lines agree with the small stellar masses determined from the
SED fits.

\subsubsection{Velocity offsets}
\label{sect:em_abs_dyn}
High-redshift galaxies show velocity offsets between \lya\ emission
and absorption lines, which are generally interpreted as galaxy scale
outflows caused by feedback from star formation. Detailed studies have
used the \lya\ line to measure the emission line redshift in
comparison to rest-frame UV absorption lines arising in the
interstellar medium (ISM) of the LBGs and found relative offsets of
several 100 km~s$^{-1}$ \citep{shapley03}, but also fainter galaxies
identified by GRBs appear to have large velocity shifts
\citep{milvang-jensen12}.  In a sample of 8 \lya\ selected galaxies
velocity offsets of 145$^{+45}_{-23}$ km s$^{-1}$ between the systemic
redshifts from rest-frame optical emission lines and the \lya\ lines
have been measured \citep{hashimoto12}. These small velocity offsets
are similar to the offsets determined from the two \lya\ emitting
galaxies in this study.  Since in the reference frame of the
outflowing gas, photons bluewards of the \lya\ line centre are shifted
into resonance, the photons escape only after having diffused to the
red wing.  \citet{verhamme06} find that \lya\ experiences about twice
the velocity shift of the true outflow velocity in low hydrogen column
density LBGs.

Some \lya\ lines have a prominent blue cutoff in the spectral profile
while others show both a blue and a much stronger red
component. Observed at lower spectral resolution, the intrinsic
redshift of the \lya\ lines is difficult to determine.  A better
approach is to compare the ISM lines with other emission lines than
\lya, preferably measured with the same instrument to avoid systematic
offsets. X-shooter spectra of the 8 o'clock arc show that the UV
absorption lines and optical emission lines have redshifts consistent
with each other \citep{dessauges11}, but since the broad ISM
absorption lines span a range of velocities relative to the systemic
one, a galaxy scale outflow of 120 km~s$^{-1}$ is likely.  When
outflows have been observed in galaxies, their ISM absorption lines
are blueshifted with respect to the systemic redshift.

Table~\ref{tab:arcz} lists both redshifts derived from emission and
absorption lines.  The absorption line redshifts are measured by
fitting Gaussian functions and are less accurate because the S/N per
pixel is low. Only for the sources with the brightest UV continuum can
we measure reliably the redshifts from absorption lines. Even though
the continuum is relatively bright, we find no significant ISM lines
in the M0359 spectrum since at $z=1.0$ most strong lines are bluewards
of the UVB spectrum.  Strong low-ionisation lines that are relatively
easily detected and are not blended include \ion{Si}{ii}
$\lambda$1260, \ion{Si}{ii} $\lambda$1304, \ion{C}{ii} $\lambda$1334,
and \ion{Si}{ii} $\lambda$1526, and among the high-ionisation species
we detect \ion{S}{iv} $\lambda\lambda$1393,1402 and \ion{C}{iv}
$\lambda\lambda$1548,1550. LBG spectra show velocity offsets between
high and low-ionisation lines \citep{shapley03}, as well as offsets
between emission and absorption lines indicative of galaxy scale
outflows. Only in M0304 do we detect several high- and low ionisation
species, but we do not detect any offset in velocity between the
lines.

To determine if velocity offsets are seen in our sample, we list in
Table~\ref{tab:phys_prop} the velocity offsets derived from the
difference between the average emission and absorption redshifts.
When the velocity offset is positive, one can interpret this as an
outflow. Within the errors, there is no evidence for large velocity
offsets in our data. A more detailed study would require higher S/N
data in particular to study the widths of the ISM lines. As the
outflows in LBGs are driven by feedback from star formation, the lower
SFRs in our galaxies relative to LBGs could result in smaller velocity
offsets. In contrast, in low-mass galaxies which have weaker
potential fields even small SFRs can drive a stronger wind. In such
cases, the specific SFR is expected to correlate with the wind
strengths as seen in $z\approx1$ galaxies \citep{kornei12}.

\subsection{Gas phase reddening}
\label{sect:ebv_gas}
For five sources, more than one of the Balmer lines are detected and
their flux ratios can be used to derive the gas phase reddening. As
the ratios of emission lines are not affected by lens magnification
and slit losses, we do not include those parameters in the
calculations. The expected emission line ratio in the absence of
reddening is a function of the gas temperatures and densities. We
adopt the values tabulated for temperatures $T$~=~10~000--20~000 K and
densities $n\sim100$ cm$^{-3}$ in \citet{brocklehurst71}. The electron
density of the gas in the lensed galaxies can be determined from the
flux ratio of the [\ion{O}{ii}] $\lambda\lambda$3727,3730 doublet. The
ratios of the lines are consistent with a low electron density $<$300
cm$^{-3}$ in all cases.

To derive the gas phase reddening, $E(B-V)_{\mathrm{gas}}$, we use all
of the Balmer line pairs available for each of the galaxies and use a
weighted average to determine the reddening. In practise this implies
that the reddening is determined by the Balmer line ratio with the
smallest uncertainty.  We assume a starburst extinction curve
\citep{calzetti00}, and list the resulting reddening in
Table~\ref{tab:phys_prop}. The reddening derived for M2031 is very
uncertain, but consistent with zero which was also found in the SED
analysis in Sect~\ref{sect:SEDfit}.

\begin{figure}
\begin{center}
\includegraphics[width=8.5cm, bb=70 360 540 710, clip]{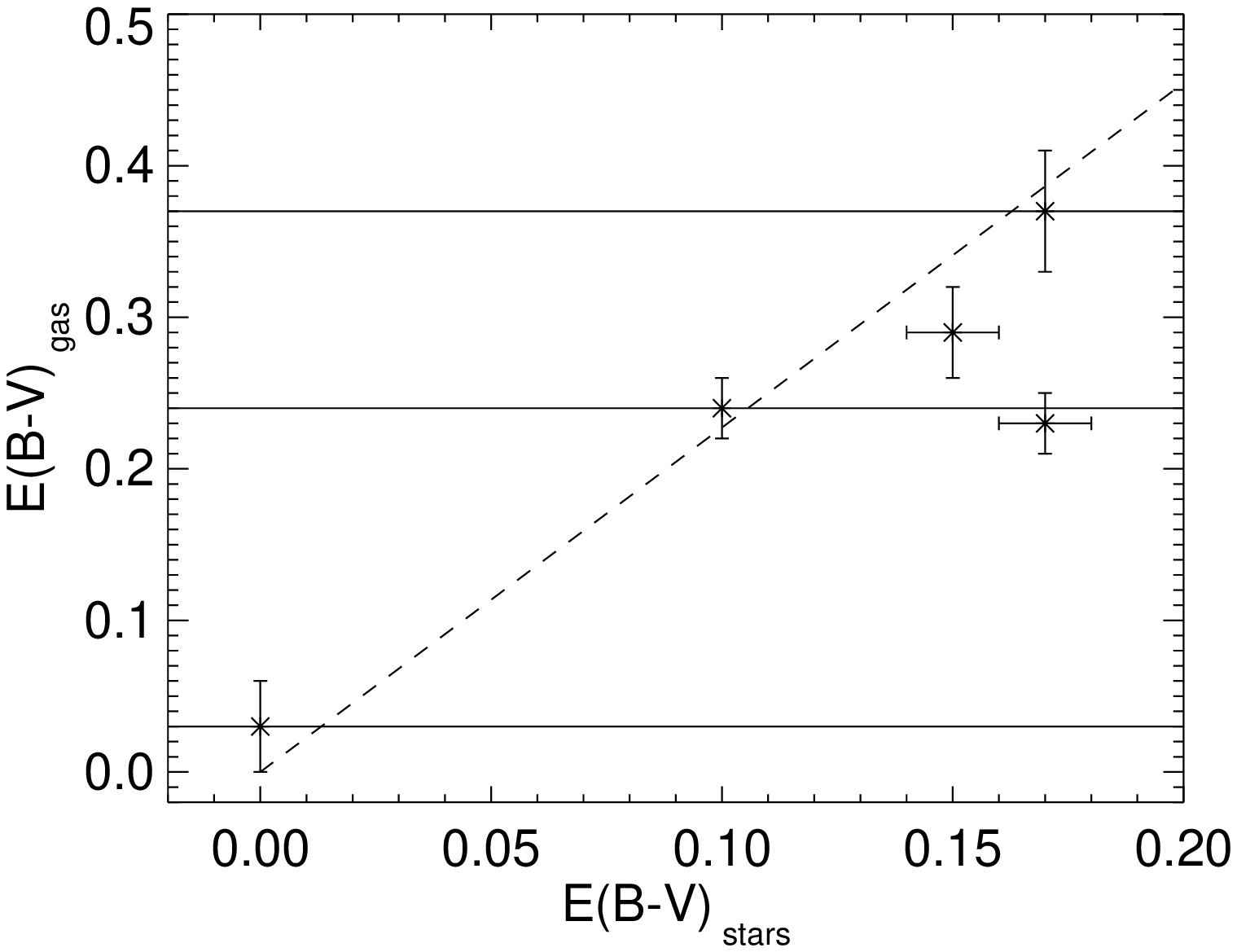}
\end{center}
\caption{Reddening in the gas phase as a function of the stellar
  reddening determined from the SED fitting. The dashed line shows
  $E(B-V)_{\mathrm{gas}}$= $E(B-V)_{\mathrm{stars}}/0.44$ from
  \citet{calzetti00}.}
\label{fig:red}
\end{figure}

In the following analyses, all the emission lines fluxes in Tables
5--13 are corrected for the intrinsic reddening
$E(B-V)_{\mathrm{gas}}$, while for those galaxies where we do not
constrain this parameter from emission line ratios we use instead the
reddening measured for the stars in Table~\ref{tab:sedfit}, and assume
a conversion to the reddening in the gas phase
$E(B-V)_{\mathrm{gas}}$= $E(B-V)_{\mathrm{stars}}/0.44$
\citep{calzetti00}. Figure ~\ref{fig:red} shows the reddening in the
gas phase as a function of the reddening of the stars determined from
SED fitting. The dashed line represents the scaling factor of 0.44
from \citet{calzetti00}, which appears to be valid for the lensed
galaxies as well.

\subsection{Starburst or AGN ionisation}
\label{sect:sb_agn}
The ionising radiation from either active galactic nuclei (AGN) or
recent star formation gives rise to strong emission lines in
galaxies. Whether the emission is caused by an AGN rather than strong
starbursts is usually investigated through emission line ratios, which
are sensitive to the hard ionisation from AGN. Conventionally, when
only the strongest rest-frame optical emission lines can be detected
in high-redshift galaxies, the line ratios \ion{[O}{iii]}/\hbeta\,
versus \ion{[N}{ii]}/\halpha\, \citep{kewley02} can be used. However,
in our sample all these strong lines are only detected in M0304 and
M0359. Both these galaxies have line ratios characteristic of star
forming galaxies.

In M2031, the \ion{C}{iv}\,$\lambda\lambda$1548,1550 lines are not
detected to a level of $3\times10^{-18}$ \ecs, so the ratio
\ion{C}{iv}/\ion{C}{iii]}$<$0.15 indicates a softer ionising spectrum,
  while an AGN would result in a fraction of 2 as argued for the Lynx
  arc \citep{binette03}. The line ratio for A31.1 is
  \ion{C}{iv}/[\ion{C}{iii},\ion{C}{iii]}~=~0.65$\pm$0.71 indicating a higher
    ionisation parameter. In AGNs, the lines \ion{N}{v}
    $\lambda\lambda$1238,1242 and \ion{He}{ii} $\lambda$1640 are
    expected to be strong, and since these lines are not detected in
    any of your spectra it suggests that the main contributor to the
    ionisation is massive stars.

 With the numerous emission lines detected from the sources it is
 possible to derive the ionisation parameter, $U$, for the
 galaxies. We use the [\ion{O}{iii}]
 $\lambda\lambda$4959,5007/[\ion{O}{ii}] $\lambda\lambda$3727,3730
 combined with the oxygen abundances to calculate log~$U$ using the
 iterative calculations in \citet{kewley02}. The resulting values lie
 in the range between --3 and --2, which is higher than measured in
 local galaxies, but typical for high redshift galaxies, both unlensed
 \citep{erb06c,erb10} and lensed ones
 \citep{hainline09,richard11,rigby11}.

\subsection{Oxygen abundance}
\label{sect:R23}

Oxygen abundances can be determined from the strong rest-frame optical
emission lines. These values are not magnification dependent, because
the parameters are derived based on line flux ratios rather than
absolute values. A more accurate method to determine abundances rely
on the detection of temperature-sensitive emission lines which are
typically faint and very rarely detected in high-redshift galaxies. In
three of the lensed galaxies at $2<z<3.5$, the temperature-sensitive
ratios of [\ion{O}{iii}] $\lambda$5007 and the \ion{O}{iii}]
  $\lambda\lambda$1661,1666 doublet allows us to use the direct $T_e$
  methods, which is presented in \citet{christensen12}.

When the temperature-sensitive lines are absent from the spectra, one
has to rely on the relations between strong emission line ratios and
the oxygen abundances calibrated from direct $T_e$ methods
\citep{osterbrock89}. The ratio defined as \( R_{23} =
(\mathrm{[OII]}\lambda\lambda3727,3730 +
\mathrm{[OIII]}\lambda\lambda4959,5007)/\hbeta\) introduced by
\citet{pagel79} is one of the most used calibrations, since the
relevant emission lines are detectable from ground-based data out to
$z=3.8$. However, the $R_{23}$ relation has two possible solutions
with a high-metallicity and a low-metallicity branch.  A wealth of
authors have published separate calibrations of the $R_{23}$
diagnostics based either on photo-ionisation models
\citep[e.g.][]{mcgaugh91,kobulnicky99,kewley02}, or direct
calibrations derived from temperature-sensitive lines
\citep[e.g.][]{alloin79,pilyugin05,nagao06}.  Typically, these
calibrations posses an intrinsic scatter of $\sim$0.2 dex, and in
absolute terms various calibrations may differ by up to 0.7 dex
relative to each other \citep{kewley08}.

A degenerate solution can be avoided if other emission lines such as
[\ion{N}{ii}] $\lambda$6586 or [\ion{Ne}{iii}] $\lambda$3869 are
detected \citep{denicolo02,kewley02,pettini04,liang06}. Also the
[\ion{O}{iii}]5007/[\ion{O}{ii}]3727 ratio can be used to discriminate
between the upper and lower branch of the calibration \citep{nagao06}.
We detect [\ion{N}{ii}] $\lambda$6586 in two of the spectra, while
[\ion{Ne}{iii}] $\lambda$3869 is detected in three galaxies. For seven
sources we can determine the $R_{23}$ ratio, including those with
direct oxygen abundance determinations \citep{christensen12}.

For the A4.1 source, the \hgamma\ flux is scaled to determine
\hbeta. Its \halpha\ to \hgamma\ ratio suggests a reddening of
\ebv\,=\,0.24$\pm$0.13, so correcting \halpha\ for the intrinsic
extinction, and given that \hbeta\ is a factor of 2.86 fainter than
\halpha\, we derive a corrected \hbeta\ flux of $(37.6\pm3.2)\times10^{-18}$
\ecs. The other emission lines are also corrected for this this
reddening value.

For some galaxies, \ion{[O}{iii]}\,$\lambda$4959 is not detected,
noisier because of telluric absorption, or close to a sky emission
line. In those cases, we use the relation $I5007=2.98\times I4959$ to
determine the total flux of the doublet \citep{storey00}.  To derive
the oxygen abundance, we use the expression for the upper and lower
branch of the $R_{23}$ calibration in \citet{pilyugin05}. 
The $R_{23}$, 12+log(O/H) values, and the propagated errors are listed
in Table~\ref{tab:phys_prop}. For completeness both the upper and
lower branch abundances are listed.  Gas phase abundances are derived
relative to the solar value of 12 + log(O/H) = 8.69 \citep{asplund09}.
We can use the additional detected lines for some of the galaxies to
constrain which of the upper or lower branch of the abundance
calibration is valid. Following \citet{cresci11}, we assume the upper
$R_{23}$ branch is valid for galaxies with
log([\ion{O}{iii}]/[\ion{O}{ii}])$<0.45$ and
log([\ion{Ne}{iii}]/[\ion{O}{ii}])$<0.6$. Whether the upper or lower
branch is preferred is indicated by the oxygen abundances listed in
bold face numbers in Table \ref{tab:phys_prop}.

To further justify the choice of the upper or lower branch of the
metallicity calibrations, Table~\ref{tab:phys_prop} also lists the
oxygen abundances using the \emph{O3N2} calibration \citep{pettini04}
and \emph{Ne3O2} calibration \citep{nagao06}. In all cases, these
calibrations agree with the preferred branch of the $R_{23}$
relation. The various oxygen abundance determinations vary within 0.3
dex, which illustrates well the known offset between various strong
line calibrations \citep{kewley08}.

In the case of M2031, M0304 and Abell 1689 ID 31.1, all have direct
abundances measurements \citep[see][]{christensen12} consistent with
those inferred from the lower branch of the $R_{23}$ calibration in
\citet{pilyugin05}.

\begin{table*}
\begin{center}
\caption{Physical properties of sources with emission lines}
\begin{tabular}{l|llllllll|}
\hline
\hline
Arc                &  A4.1         & A31.1  &  M0304  &  M0359 &   M0416      & 
   M1311    & M2031  & Section\\
\hline

Velocity dispersion $\sigma$ (km s$^{-1}$) & 17$\pm$1 & 22$\pm$2 &
56$\pm$1& 53$\pm$2 & 65$\pm$9 & 53$\pm$3 & 43$\pm$1 &
\ref{sect:dynmass}\\

\smallskip

$\Delta V$ ($z_{\mathrm{em}}-z_{\mathrm{abs}}$) (km s$^{-1}$) & --- & 95$\pm$131 & --10$\pm$55 &
--- & --- & --- & 80$\pm$90  & \ref{sect:em_abs_dyn}\\

SFR$(\mathrm{[OII]})$ (M$_{\odot}$ yr$^{-1}$)  & 0.7$\pm$0.1  &
1.8$\pm$0.3 & 55$\pm$11 & 8.1$\pm$2.7 &  4.7$\pm$0.7 & 3.0$\pm$1.8 & 7.1$\pm$3.3 & \ref{sect:sfr} \\

SFR$(\mathrm{\halpha})$ & 0.4$\pm$0.1 & --- & 16$\pm$3 &
1.5$\pm$0.5 & ---& --- & 14.5$\pm$6.5 \\

\ebv$_{\mathrm{gas}}$ (mag)   & 0.24$\pm$0.13 & 0.37$\pm$0.28 &
0.23$\pm$0.01 & 0.29$\pm$0.01 & ---&--- &
0.03$\pm$0.33 & \ref{sect:ebv_gas} \\

log~$U$  & --2.7 &  --2.1 & --2.8 & --3.1 & --2.4 & --3.1 &
--2.1 & \ref{sect:sb_agn} \\

\hline

\smallskip

log $R_{23}$      & 0.81$\pm$0.04 & 0.85$\pm$0.05 & 0.94$\pm$0.001 &
0.67$\pm$0.01 & 0.81$\pm$0.04 & 0.19$\pm$0.06 & 0.91$\pm$0.02 & \ref{sect:R23} \\

12+log(O/H) (upper branch)& \bf{8.33$\pm$0.08} &  8.45$\pm$0.04 &
8.22$\pm$0.01 & {\bf 8.18$\pm$0.05} & {\bf 8.44$\pm$0.06} & {\bf 8.66$\pm$0.14} & 8.39$\pm$0.01 & \ref{sect:R23}\\

\smallskip

12+log(O/H) (lower branch)& 7.91$\pm$0.11 & {\bf 7.63$\pm$0.10} & {\bf
  8.16$\pm$0.01} & 7.48$\pm$0.04 &  7.73$\pm$0.11 & 7.13$\pm$0.01 &
{\bf 7.74$\pm$0.03}  &  \\

\smallskip

$Z/Z_{\odot}$ (lower; upper)  & 0.2; {\bf 0.4} & {\bf 0.1}; 0.6 &
{\bf 0.3}; 0.6  &
0.3; {\bf 0.3}  & 0.1; {\bf 0.3} & 0.03; {\bf 0.9} & {\bf 0.1}; 0.5 &   \\

12+log(O/H) (\emph{O3N2}) &  ---   &  ---& 8.04$\pm$0.01 &
8.48$\pm$0.01 & --- & --- & --- \\

12+log(O/H) (\emph{Ne3O2}) &   8.24$\pm$0.07  &  7.85$\pm$0.15 & 8.20$\pm$0.01 &--- & ---&--- &
7.56$\pm$0.11 \\

\hline
\end{tabular}
\label{tab:phys_prop}
\end{center}

\flushleft{
\textit{Notes.~} \\
The listed SFRs assume a Chabrier IMF as described in the text.
For completeness we list both the upper and the lower branch
calibrations for the $R_{23}$ diagnostics. The values in bold face
represent the preferred values based on the ratios of either
[\ion{Ne}{iii}]/[\ion{O}{ii}] or [\ion{O}{iii}]/[\ion{O}{ii}].
}

\end{table*}

\subsection{Star formation rates}
\label{sect:sfr}
We determine the SFRs using three methods: [\ion{O}{ii}] and
\halpha\ emission lines, and the UV continuum flux at rest-frame
1500--2800~{\AA}.  The measured emission line fluxes are corrected for
the slit losses and magnification factors (Table ~\ref{tab:slitloss}),
and we also correct the fluxes for the intrinsic reddening.  From the
sum of the flux in the doublet [\ion{O}{ii}] $\lambda$3727,3730 we
calculate the extinction-corrected luminosity
($L_{\mathrm{[OII],cor}}$ based on the gas phase reddening,
$E(B-V)_{\mathrm{gas}}$), and use the relation in \citet{kennicutt98}
to derive the intrinsic SFR. We divide the conversion factor by $1.7$
to correct the SFR from a Salpeter IMF to a Chabrier IMF. In a similar
manner we calculate the SFR based on the \halpha\ lines, which is
detected for three of the galaxies. Again the SFR conversion is
divided by a factor of $1.7$.  The resulting values are listed in
Table~\ref{tab:phys_prop}. In addition, we also calculate the SFR from
M2031 assuming that \halpha\ is 2.86 times the \hbeta\ flux in case of
a zero reddening.  In the following section~\ref{sect:fmr}, we use the
SFRs based on the [\ion{O}{ii}] $\lambda\lambda$3727,3730 doublet.
Once corrected for reddening, SFRs based on [\ion{O}{ii}] and
\halpha\ are generally in good agreement \citep[e.g.][]{kewley04}, and
\halpha\ is known as a good tracer of the current ongoing SFR.

As an alternative, we also calculate the SFR from the rest-frame UV
luminosity at 1500--2800 {\AA} using the calibration in
\citet{kennicutt98}, again correcting the scaling relation for the IMF
difference by dividing by a factor $1.7$. We make use of the best fit
spectral models available for all of the galaxies in our sample,
correct these for the reddening $E(B-V)_{\mathrm{stars}}$, and
determine the UV flux at the rest-frame 1500 {\AA}. The resulting SFRs
are listed in Table \ref{tab:sedfit}.

The uncertainties for the SFRs are propagated and contain the
uncertainties measured either for the [\ion{O}{ii}] flux or the UV
continuum as well as the uncertainty for the slit loss and
magnification factors. When two or more estimates of the SFRs can be
derived, we find that the values are consistent to within a factor of
two. In contrast to the SFRs, the specific SFRs, i.e. the SFRs divided
by the stellar mass, do not depend on the magnification factor, slit
losses or choice of IMF. The inverse of the SSFRs can be taken as the
formation time-scale of the galaxy given that the SFR is constant.
For a galaxy with a small SFR and high stellar masses, the
correspondingly low SSFR suggests that the SFR must have been higher
in the past. In particular, M2129, the oldest galaxy in our sample,
shows evidence of only little present ongoing star formation. The
[\ion{O}{ii}] $\lambda\lambda$3727,3730 doublet is detected at a level
corresponding to a SFR~=~0.8 M$_{\odot}$ yr$^{-1}$, implying that much
stronger star formation has taken place in the past to build up the
stellar mass in this massive galaxy.

\section{Fundamental relation for star-forming galaxies}
\label{sect:fmr}

\begin{figure}
\begin{center}
\includegraphics[width=8.5cm, bb=80 370 520 700, clip]{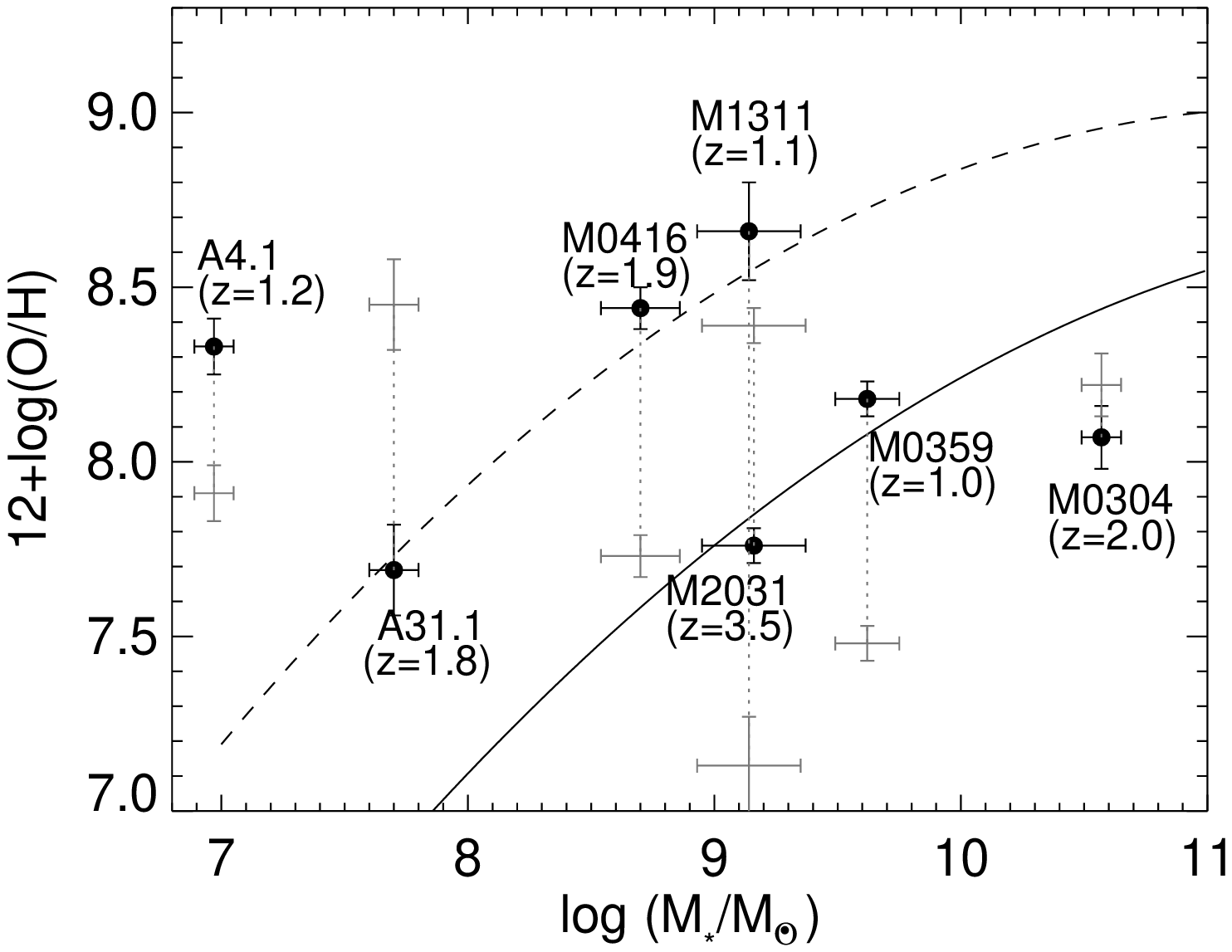}
\end{center}
\caption{Oxygen abundances of the lensed galaxies as a function of the
  stellar masses, corrected for lens magnifications. The black circles
  show the oxygen abundances from the preferred branch of the $R_{23}$
  relation including the direct measurements for A31.1, M0304 and
  M2031, while the grey error-bars show the location of the galaxies
  if the alternate branch of the $R_{23}$ relation is adopted for all
  galaxies.  The mass-metallicity relation for Lyman break galaxies at
  $z=3-4$ is shown by the solid line \citep{maiolino08,mannucci09},
  and galaxies at $z=0.7$ by the dashed line \citep{savaglio05}.}
\label{fig:mass-metal}
\end{figure}

\subsection{Mass-metallicity relations}
Observations of mass-metallicity relations have revealed an evolution
with increasing redshifts from local galaxies in SDSS
\citep{tremonti04}, field galaxies at $z=0.7$ \citep{savaglio05}, and
out to $z>3$ \citep{maiolino08,mannucci09}. \citet{richard11} analyse
a larger sample of lensed galaxies at $1.5<z<3.3$, and find the galaxy
metallicities at $z\sim3$ are slightly larger than field galaxies.

Figure~\ref{fig:mass-metal} shows that the lensed galaxies exhibit a
huge scatter compared to the established relations, and we do not find
any clear correlation. The highest stellar mass galaxy, M0304, belongs
to a merging system, and therefore naturally has a lower observed
metallicity relative to the expected value from extrapolation of the
mass-metallicity relations.  

 Since the mass-metallicity relations are based on various emission
 line diagnostics, which exhibit offsets with respect to each other
 \citep{kewley08}, it is not straight forward to analyse the evolution
 of the mass-metallicity relation with redshift when different galaxy
 samples are involved. Such offsets in abundance diagnostics can
 partly explain the reason why the lensed galaxies do not follow any
 of the correlations at a given redshift.

\subsection{Fundamental relation at low stellar masses}

 The fundamental relation between galaxies SFRs, oxygen abundances,
 and stellar masses calibrated for low-redshift, high-mass galaxies
 \citep{lara-lopez10,mannucci10} have revealed a larger scatter when
 extending relation to lower stellar masses.  In particular, GRB host
 galaxies with $M_*\lesssim10^9$ M$_{\odot}$ show a larger scatter
 relative to the fundamental relation \citep{kocevski11,mannucci11},
 and \citet{mannucci11} provide an extension of the fundamental
 relation at lower stellar masses. Recent observations show that also
 a low stellar mass supernova Type Ia host galaxy at $z=1.55$ follows
 this relation \citep{frederiksen12}.

\begin{figure}
\includegraphics[width=8.5cm, bb=80 370 520 710, clip]{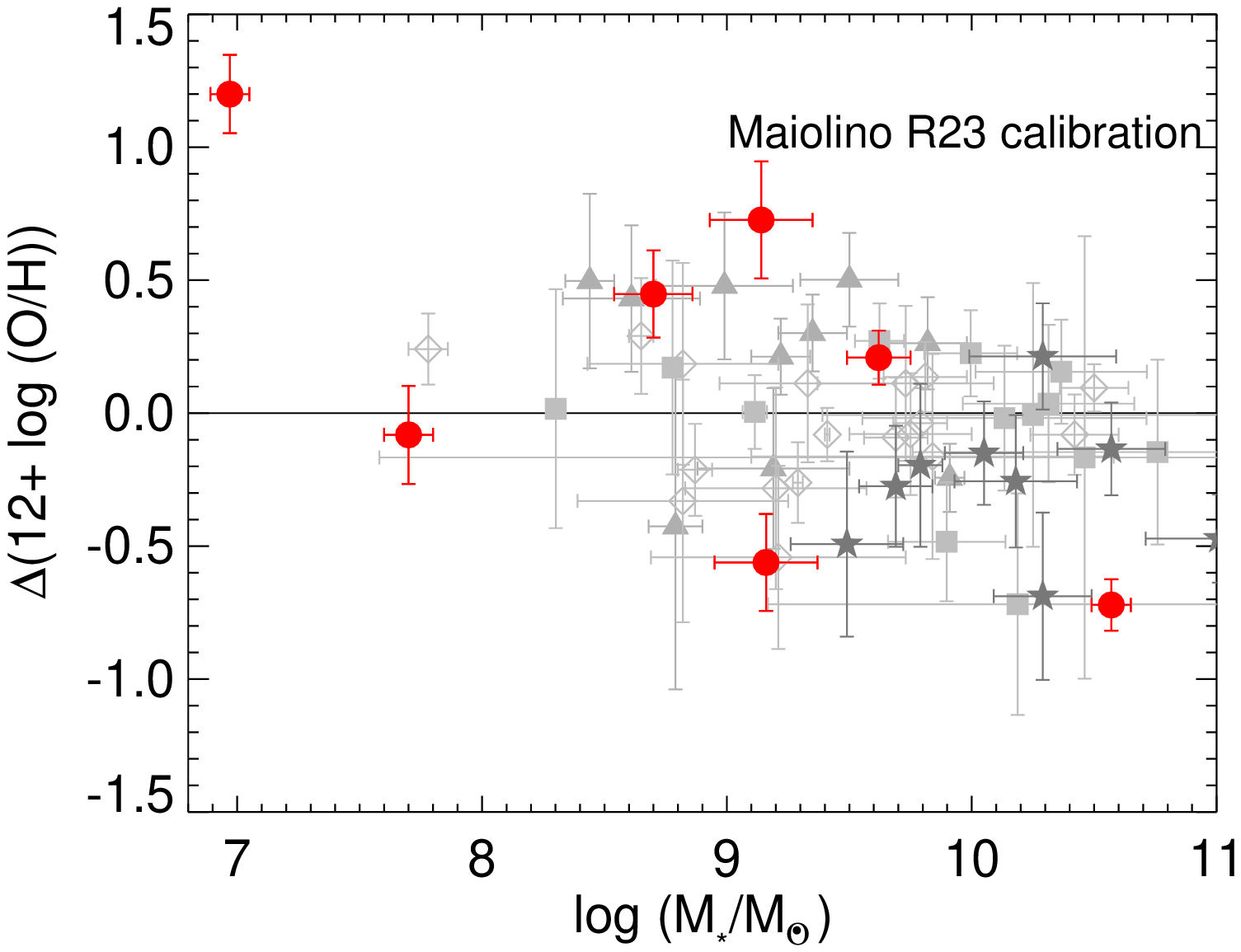}
\caption{Offsets of the derived oxygen abundance compared to the
  predicted one from the fundamental relation in \citet{mannucci11}.
  The red circles show the location of the preferred $R_{23}$ branch
  with the calibration in \citet{maiolino08}, and using the direct
  measurements for A31.1, M0304 and M2031 in \citet{christensen12}. A
  location above zero implies that the measured metallicity is higher
  than expected from the relation. The grey squares show the offsets
  of 13 lensed galaxies at $z=2-3$ for which either the $R_{23}$ or
  the \emph{N2} calibration from \citet{maiolino08} were used
  \citep{richard11}. The grey triangles show offsets of 10 lensed
  galaxies at $z=1-2$ in \citet{wuyts12} for which the oxygen
  abundance was derived from the \emph{N2} calibration in
  \citet{maiolino08}. The outlined diamonds represent 18 GRB hosts at
  $z<1$ \citep{mannucci11}, and the star symbols represent unlensed
  and more massive $z\sim3.5$ galaxies \citep{maiolino08}.}
\label{fig:fp}
\end{figure}

If we calculate the differences between the oxygen abundances in
Table~\ref{tab:phys_prop} and the calibration in \citet{mannucci11},
we find that the galaxies exhibit a large scatter compared to the
relation. A potential risk when analysing and comparing abundances, is
that there is a significant offset between various emission line
diagnostics \citep{kewley08}. We therefore also derive the oxygen
abundances using the calibrations in \citet{maiolino08}, which
correspond to the abundances used in the derivation of the fundamental
relation in \citet{mannucci10}. The offsets are illustrated in
Figure~\ref{fig:fp}.  Other low-mass lensed galaxies at $z=1-3$
\citep{richard11,wuyts12} illustrated by squares and triangles in
Fig.~\ref{fig:fp} show a similar trend of not following the
fundamental relation. The standard deviation of the offsets from the
relation is 1 dex, and the galaxy with the lowest stellar mass, A4.1,
is offset by $\sim$1 dex. In any case, the high abundance of this
galaxy is unexpected given its low stellar masses. The basis of the
selection of the upper branch for the $R_{23}$ abundance calibration
is that the flux ratios of [\ion{Ne}{iii}]/[\ion{O}{ii}] and
[\ion{O}{iii}]/[\ion{O}{ii}] suggest a higher abundance
\citep{nagao06,cresci11}.

In addition to the large scatter, the locations of the galaxies have a
slope of --0.3 per dex in log $M_*$, which is statistically
significant with a Spearman nonparametric correlation test giving a
probability of 0.98. The calibration of the fundamental relation in
\citet{lara-lopez10} also produces a slope and a large scatter.

Our finding of a discrepant location of the lensed galaxies compared
to the fundamental relation for star-forming galaxies in
\citet{mannucci10} and the extension to low-mass galaxies in
\citet{mannucci11} does not imply that a fundamental relation is not
present. It merely suggests that the scatter is significant when
low-mass galaxies are analysed.  Massive lensed galaxies analysed in
the literature follow the fundamental relation when their
metallicities are 12+log(O/H)~$>$~8.0, and stellar masses larger than
$10^{10}$ M$_{\odot}$ \citep{dessauges11,richard11}, while lensed
galaxies with stellar masses below $\sim10^{10}$ M$_{\odot}$ have in
majority larger metallicity than expected from the fundamental
relation \citep{richard11,wuyts12} consistent with our findings.

If we assume that the oxygen abundance in the fundamental relation
equation depends only on the SFR, the galaxies would fall within 0.5
dex of a relation. Similarly, a smaller scatter with respect to the
fundamental relation can be obtained if the SFR is not included, and
only the stellar mass is a dependent variable. However, the minimum
scatter with respect to the relation is maintained with a combination
of dependence of both the stellar mass and the SFR. \citet{mannucci10}
defines a projection of the fundamental relation: \(\mu_{\alpha} =
\log(M_*) -\alpha \log(\mathrm{SFR})\), where $\alpha=0.32$ gives the
smallest scatter. For the lensed galaxies with stellar masses below
$10^{10}$ M$_{\odot}$, we determine the smallest scatter at
$\alpha=0.36$.

\begin{figure}
\begin{center}
\includegraphics[width=8.5cm, bb=80 370 530 710, clip]{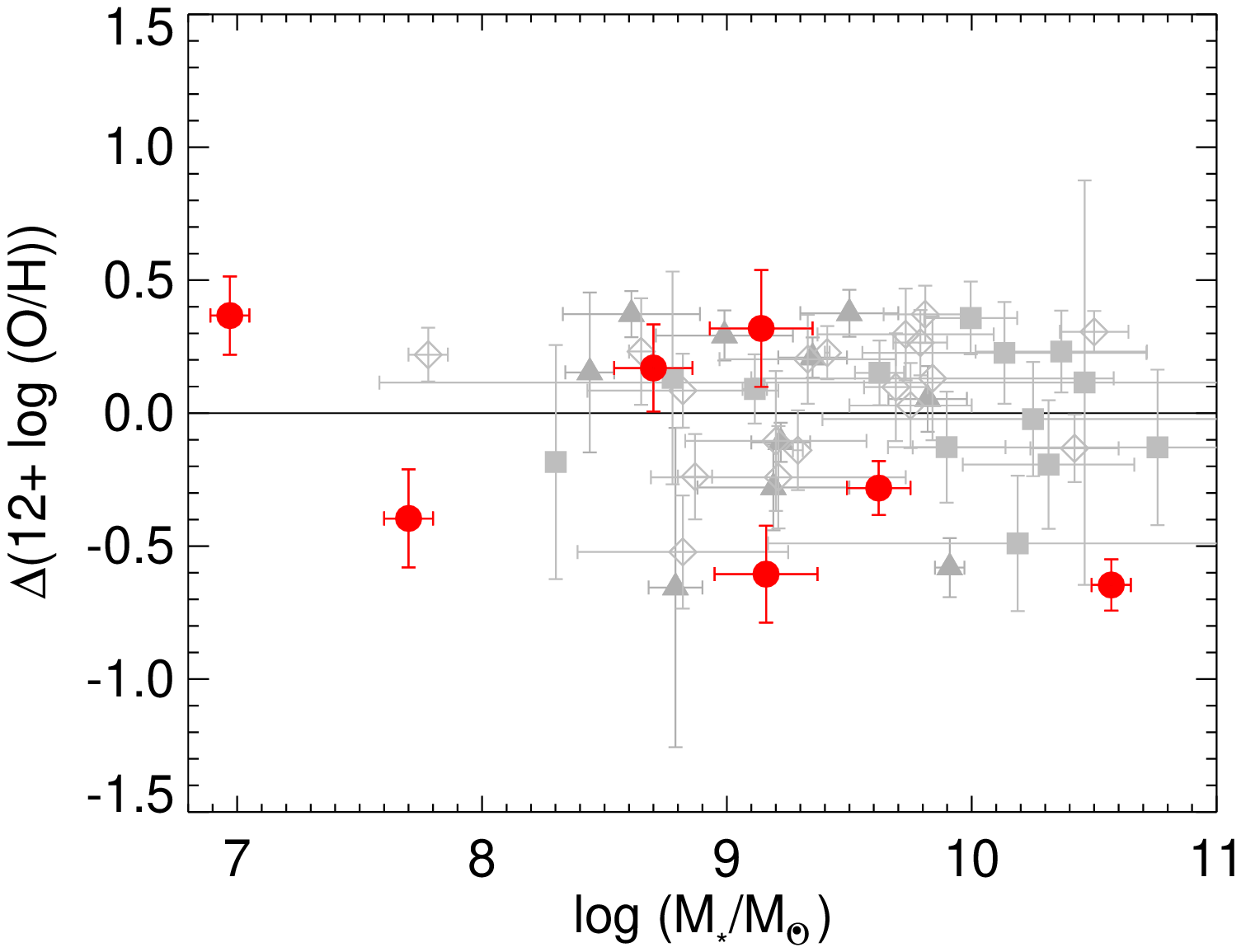}
\end{center}
\caption{Offset in measured abundance compared to the new calibration
  of the fundamental relation given by Eq. (1). The symbols present
  the same data as in Fig.~\ref{fig:fp}.  From the entire sample of
  lensed galaxies combined with low redshift GRB hosts, we measure a
  standard deviation of $\sigma=0.3$ dex.}
\label{fig:newfp}
\end{figure}

Since the sample of lensed galaxies is so small compared to the
140\,000 SDSS galaxies, we can only suggest an alternative fit for the
fundamental relation valid for low mass galaxies. The oxygen
abundances for all the lensed galaxies in the right panel of
Fig.~\ref{fig:fp} and the GRB hosts from \citet{mannucci11} are fit
with the functional form of the fundamental relation in
\citet{mannucci10}.  The resulting fit is:
\begin{eqnarray}\nonumber
\mathrm{12+log(O/H)} & = & 8.52 + 0.26m + 0.013s + 0.025m^2\\
&&-0.026ms + 0.013s^2,
\end{eqnarray}
where $s$~=~log(SFR) and $m$~=~log($M_*$) -- 10 in solar units. With
this alternative calibration, the scatter is 0.3 dex and the slope is
absent as demonstrated in Fig. ~\ref{fig:newfp}.  A Spearman test
gives a probability of 0.3 for a correlation.

One concern is that the $R_{23}$ calibration is not an accurate
metallicity tracer given the large offsets between various strong line
calibrations \citep{kewley08}. A comparison of the oxygen abundances
from $R_{23}$ and the direct $T_e$ method for metal-rich, local SDSS
galaxies suggests that the $T_e$-based log(O/H) abundances are lower
by up to 0.6~dex compared to the upper branch of the $R_{23}$ relation
\citep{liang07}.  However, for the galaxies in our sample we find that
the $R_{23}$ calibration in \citet{pilyugin05} is in agreement with
direct $T_e$ based abundances for three of the galaxies
\citep{christensen12}. For the other four galaxies we can only assume
that this is valid too.

Since the lens magnification as well as the slit loss correction
affect both the SFR and stellar masses equally, any uncertainty in the
magnification factor and correction factor has little impact on the
location of the galaxies relative to the fundamental plane. The only
assumption is that the emission lines are uniformly distributed over
the entire galaxy, such that the emission line flux follows the
continuum flux. Metallicity gradients are known to be present in disk
galaxies, so we likely overestimate the integrated oxygen abundance
slightly. Considering that gradients are typically a few times 0.1 dex
over the face of a large (local) galaxy, and that the lensed galaxies
in this study are compact sources dominated by at most a few
luminous \ion{H}{ii} regions, any metallicity gradient cannot explain
an integrated 1 dex offset relative to the fundamental relation.

\section{Discussion and summary}
\label{sect:conclusions}

We have presented spectroscopic data and determined redshifts for 12
strongly lensed galaxies behind various massive galaxy clusters. Three
of the redshifts were previously determined by other studies.
Although the selection of the lensed galaxies is based on optical
data, and therefore makes a preferential selection of UV bright
high-redshift galaxies, we still achieve a sample of various galaxy
properties (young and old, reddened and non-reddened, star-forming and
non-star-forming), because we do not include colour selections.  The
lensing magnification effect allows us to carry out detailed
investigations of galaxies that are on the average 1--2 orders of
magnitude fainter and less massive than galaxies studied in unlensed
field populations. The galaxies are also intrinsically fainter than
other lensed galaxies studied in the literature, simply because we
targeted galaxies with observed magnitudes $\sim$24, which is fainter
than the ones investigated previously.

Using the stellar masses, SFRs and oxygen abundances, we find a
breakdown of the fundamental relation for star-forming galaxies at
very low stellar masses ($M<10^{10} M_{\odot}$) with offsets of up to
1--2 dex, which implies that one cannot use the calibration in
\citet{mannucci10} to derive any missing parameter (SFR, $M_*$, or
oxygen abundance) when two other parameters are known for
high-redshift galaxies. Contrary to the suggestion that the relation
is not valid at $z>2.5$, we find one galaxy at $z=3.5$ with a
metallicity only 0.4$\pm$0.1 dex too low compared to the expectation
from the fundamental relation. We provide an alternative calibration
extending the low-mass end of the fundamental relation.

Our findings of the SFRs from various indicators (UV continuum versus
either [\ion{O}{ii}] or \halpha\ emission lines) agree with each other
within a factor of two after correcting for the intrinsic
reddening. Such spreads in the SFR estimates from various diagnostics
are known for LBGs \citep{erb03}. However, \citet{wuyts11} find that
the SFRs after dust correction in four lensed $z=2$ galaxies agree
under the assumption that the extinction law is steeper than a
Calzetti law as observed for unlensed LBGs \citep{reddy10}. In the
lensed galaxies studied here, the UV-based SFRs are on the average
lower than inferred from emission lines, which would support the
suggestions of a steeper extinction curve, such that the UV flux
should be corrected by a larger factor to match the SFRs based on
emission lines.  However, even if the SFRs are overestimated by a
factor of two, this would make no impact on the offset of the lensed
galaxies with respect to the fundamental relations between the SFR,
stellar mass and oxygen abundances explored in Sect.~\ref{sect:fmr}.

We find no evidence for velocity offsets between rest-frame UV and
optical emission lines and UV absorption lines when both are detected,
since the offsets we derive are only significant on the 1$\sigma$
level. Observations of high-redshift galaxies have interpreted offsets
between \lya\ and UV absorption lines or optical emission lines as
indication of galaxy outflows of up to several 100 km s$^{-1}$, even
though the velocity offset of \lya\ emission lines can be explained by
radiation transfer effect in expanding neutral gas without the need
for extreme velocities.  The combination of accurate redshift
determination from intermediate resolution spectra combined with a
full spectral coverage is essential to investigate individual galaxies
for these kinematic effects. High-velocity outflows from galaxies can
also be inferred from the extension in velocity space of the ISM
absorption lines, but in the low S/N ratio spectra investigated here,
we cannot determine if the absorption lines span a large velocity
range.

One of the driving questions behind this investigation is which
galaxies have strong emission lines that are detected with shallow
spectra. Galaxies without bright knots (or equivalent to \ion{H}{ii}
regions), do not have strong emission lines. A8.1 and M1115 show no
emission lines at all, while M1311 has only very faint lines. Abell
1689 ID 8.1 is a giant arc, which has a blue continuum suggesting a
young galaxy with ongoing star formation.  With a redshift $z=2.68$,
it should have detectable emission lines, also beyond \lya. The
\lya\ line itself could easily be absorbed as it is only detected in
about 25\% of LBGs. On the other hand, other lines could be present
but just absorbed or obscured. As seen in the M1206 arc, we only
detect a single but very strong emission line from [\ion{O}{iii}]
$\lambda$5007, while all the other strong emission lines in the large
wavelength range spanned by X-shooter fall behind strong sky emission
lines, or absorbed by telluric lines. In the case of M1115, which is a
low-surface brightness arc, the photometric redshift is either low or
high, so emission lines are expected. However, the SED of the galaxy
suggests an old stellar population, where emission lines in
\ion{H}{ii} regions are no longer produced.

Alternatively, we can compare the relative luminosities at the
rest-frame UV and optical. Not surprisingly, the galaxies with low
absolute UV luminosity relative to that in the rest-frame (M1115 and
M2129) do not exhibit strong emission lines. When comparing the
rest-frame UV and optical luminosities in Table~\ref{tab:mag_r}, most
galaxies can be separated into UV bright or UV faint relative to their
optical luminosities. We find that the galaxies without emission lines
are those with fractional luminosities $L/L^*$(UV)$\lesssim
L/L^*$(optical). The only exception is again A8.1.

In future studies when large numbers of gravitational lensed
candidates are detected in multi-band images covering large areas on
the sky, we intend to focus not only on galaxies with obvious clumpy
morphologies indicating the presence of \ion{H}{ii} regions and
on-going star formation, but also on the galaxies with steep spectral
slopes.

\section*{Acknowledgments}
The Dark Cosmology Centre is funded by the DNRF.  LC acknowledges the
support of the EU under a Marie Curie Intra-European Fellowship,
contract PIEF-GA-2010-274117. JR is supported by the Marie Curie
Career Integration Grant 294074.  PL acknowledges funding from the
Villum foundation. BMJ acknowledged support from the ERC-StG grant
EGGS-278202. We thank Daniel Schaerer for discussions about HyperZ,
and Roser Pell\'o for modifying HyperZ to allow the fitting of
spectra. Thanks to Bram Venemans for discussions about X-shooter data
reduction, and Stefano Covino, Valerio D'Elia, Johan P.~U. Fynbo,
Daniele Malesani, Hans Ulrik N{\o}rgaard-Nielsen, and Beate Stelzer
for carrying out the observations. We also thank the anonymous referee
for comments and suggestions that improved the presentation of the
results.

\bibliographystyle{apj}
\bibliography{ms_lc}
\end{document}